\documentclass[prd,aps,floats,twocolumn,nofootinbib]{revtex4}
\usepackage{commath}
\pdfoutput=1
\usepackage[usenames, dvipsnames]{color}
\usepackage{graphicx, array}
\usepackage{multirow}
\usepackage{graphics}
\usepackage{graphics,epsfig}
\usepackage{amssymb}
\usepackage{amsmath}
\usepackage{ulem}
\usepackage{multirow}
\usepackage{subfigure}
\usepackage{bm}
\usepackage{txfonts}
\usepackage{cancel}
\usepackage{url}
\usepackage{hyperref}

\usepackage{ulem}
\newcolumntype{C}[1]{>{\centering\arraybackslash}m{#1}}

\def\be{\begin{equation}}
\def\ee{\end{equation}}
\def\bi{\begin{itemize}}
\def\ei{\end{itemize}}
\def\ben{\begin{enumerate}}
\def\een{\end{enumerate}}
\def\bt{\begin{tabular}}
\def\et{\end{tabular}}
\def\bc{\begin{center}}
\def\ec{\end{center}}

\def\bea{\begin{eqnarray}}
\def\eea{\end{eqnarray}}

\def\ba{\begin{eqnarray}}
\def\ea{\end{eqnarray}}
\def\Rtype{$R$-type~}
\def\Stype{$S$-type~}
\def\Rtypes{$R$-types~}

\def\phiG{\phi_{5}(\tilde{r},{\tilde{R}})}

\let\oldhat\hat
\renewcommand{\vec}[1]{\boldsymbol{\mathbf{#1}}}
\renewcommand{\hat}[1]{\oldhat{\boldsymbol{\mathbf{#1}}}}

\graphicspath{ {./images/},{ ./images/ratioFigs/},{ ./images/windowFigs/},{ ./Version1/}}

\begin{document}

\input{epsf}

\title{Testing $f(R)$ Gravity With Scale Dependent Cosmic Void Velocity Profiles}
\author{Christopher Wilson and Rachel Bean}

\affiliation{Department of Physics, Cornell University, Ithaca, New York 14853, USA.}
\affiliation{Department of Astronomy, Cornell University, Ithaca, New York 14853, USA.}

\begin{abstract}

We study the impact of cosmological scale modifications to general relativity on the dynamics of halos within voids by comparing  N-body simulations incorporating Hu-Sawicki $f(R)$ gravity, with $|f_{R0}|=10^{-6}$ and $10^{-5}$, to those of $\Lambda$CDM. By examining the radial velocity statistics within voids classified based on their size and density-profile,  as ``rising"  ($R$-type) or ``shell" ($S$-type), we find that halo motions in small \Rtype voids, with effective radius $<15 Mpc/h$, reveal distinctive differences between $f(R)$  and $\Lambda$CDM  cosmologies.

To understand this observed effect, we study the linear and nonlinear fifth forces, and develop an iterative algorithm to accurately solve the nonlinear fifth force equation. We use this to characterize the Chameleon screening mechanism in voids and contrast the behavior with that observed in gravitationally collapsed objects.  

The force analysis underscores how smaller \Rtype voids exhibit the highest ratios of fifth force to Newtonian force, which source  distinguishable differences in the velocity profiles and thereby provide rich environments in which to constrain gravity.

\end{abstract}

\maketitle

\section{Introduction}
The observed late time acceleration of the universe \citep{Perlmutter:1998np, Riess:2004nr} has been shown through a broad set of cosmological observations to be consistent with the inclusion of a cosmological constant term $\Lambda$ in the Einstein equations, equivalent to introducing a form of dark energy  \citep{Eisenstein:2005su, Percival:2007yw, Percival:2009xn, Kazin:2014qga, Spergel:2013tha, Ade:2013zuv, Ade:2015xua}. When comparing observational values of $\Lambda$ to predictions from high energy physics, one finds a mismatch of $\Lambda_{obs}/\Lambda_{theory} \simeq 10^{-120}$, motivating a search for alternative theories  to $\Lambda CDM$, including those which induce a deviation from general relativity (GR) on cosmological scales $\sim1/\Lambda_{obs}$.

The landscape of modified theories of gravity is extremely broad \citep{Clifton:2011jh}. A feature shared across many of them is a new scalar degree of freedom which mediates the ``fifth force" and parametrizes deviations from GR. Due to observational constraints, any viable theory of gravity which modifies GR on cosmic scales to account for the late time acceleration, must also have a mechanism to ``screen" the fifth force in solar systemlike environments, to reduce to GR and pass local tests of gravity. 
Theories employing the chameleon mechanism \citep{Khoury:2003rn} feature a scalar field nonminimally coupled to matter such that the mass of the field becomes large in regions of high density, thereby suppressing the fifth force. The most popular class of such models  is $f(R)$ gravity, which modifies $GR$ by replacing the Einstein-Hilbert action with a general function of the Ricci Scalar $f(R)$. Hu and Sawicki  \citep{Hu:2007nk} demonstrated that this function can be chosen to match a $\Lambda$CDM cosmology without the need to include dark energy, making it a viable alternative to GR. By conformally transforming the metric, it can be shown that $f(R)$ gravity is equivalent to GR plus a nonminimally coupled scalar field which undergoes Chameleon screening (see \cite{Sotiriou_2010} or \cite{Nojiri_2011} for a review). An alternative screening mechanism is provided by the Vainshtein mechanism \citep{VAINSHTEIN1972393}, seen in ``braneworld" theories of gravity such as nDGP \citep{Dvali_2000}. Here the scalar mediating the fifth force is screened whenever its derivatives grow large, such as in the vicinity of sizable overdensities, see for example \cite{Brax:2012gr}. 

 Voids by definition are underdense regions of the cosmic web \cite{1978ApJ...222..784G}, where due to the low density, potential modifications to gravity may become unscreened and lead to observational differences from GR.  There has been a wealth of research using cosmological simulations that incorporate the effects of modified gravity theories to study void statistics in $f(R)$ \citep{Li:2011pj,Zivick:2014uva,Perico:2019obq,Contarini:2020fdu,Padilla:2014hea,Cai:2014fma,Davies:2019yif}, $nDGP$ \citep{Falck:2017rvl,Paillas:2018wxs}, and Galileon \citep{Baker:2018mnu,Barreira:2015vra} gravity scenarios. Voids have also been shown to provide a rich environment to investigate  dark energy through multiple observable quantities. This includes void number counts as function a of size \citep{Sheth2004AHO,Pisani:2015jha,Wojtak:2016brz,Adermann:2017izw,Contarini:2019qwf}, void density profiles (void-halo correlation function) \citep{Ceccarelli:2013rza,Ricciardelli:2014vga,Novosyadlyj:2016nnu,Massara:2018dqb,2020PhRvL.124v1301N} and void dynamics and velocity profiles \citep{AragonCalvo:2012bd,Lambas:2015afa}.  The impact of voids on weak gravitational lensing \citep{Krause:2012aq,Chantavat:2014gqa,Cai:2016rdk,Davies:2018jpm,Davies:2020udw,2020ApJ...890..168R}, redshift space distortions and gravitational redshift effects \citep{Hamaus:2015yza,Hamaus:2016wka,Cai:2016jek,2019MNRAS.483.3472N,Chuang:2016wqb,Sakuma:2017hfc,2019PhRvD.100b3504N,Correa:2020ddy,2020MNRAS.499.4140N}, the integrated Sachs-Wolfe effect \citep{Nadathur:2011iu,2016ApJ...830L..19N}, and the kinetic Sunyaev-Zel'dovich effect \citep{Li:2020bsu} have also been studied.

Recent galaxy and CMB surveys have demonstrated how observational data from voids can provide cosmological constraints. The Sloan Digital Sky Survey (SDSS) has provided a wealth of observational data including void density profiles \citep{2015MNRAS.449.3997N}, void lensing profiles \citep{Clampitt:2014gpa}, redshift space distortions around voids \citep{Achitouv:2019xto,Hamaus:2020cbu,Aubert:2020lfr}. The Dark Energy Survey (DES) data has been used to study weak gravitational lensing around voids \citep{Sanchez:2016mky,Fang:2019xti}, and to combine DES-detected voids to derive Planck CMB void lensing signatures \citep{Vielzeuf:2019awz}. Upcoming spectroscopic and photometric experiments, such as DESI \citep{Levi:2013gra}, Euclid, the Nancy Grace Roman Space Telescope (previously WFIRST) \citep{2019arXiv190205569A,Eifler:2020vvg} and the Rubin Observatory LSST survey  \citep{Abell:2009aa,Abate:2012za}, will provide new opportunities to further probe gravity on large scales within void environments.  

The paper is structured as follows: Section \ref{sec:formalism} lays out the formalism used in this paper -- including the modified gravity modeling in Sec.~\ref{sec:MG}, the cosmological simulations utilized in Sec.~\ref{sec:sims}, and the void identification and classification scheme in Sec.~\ref{sec:voidident}. In Sec.~\ref{sec:results} we present the main findings of the paper -- summarizing the effects of modified gravity on void density profiles in Sec.~\ref{sec:densityprofiles}, and the impact on halo radial velocity profiles within the voids in Sec.~ref{sec:radvel}. The findings are analyzed in Sec.~\ref{sec:analysis} -- discussing the impacts of linear and nonlinear estimates of the fifth force in Sec.~\ref{sec:linfifth} and \ref{sec:nonlinfifth} respectively, and how screening behaves in voids in \ref{sec:screening}. In Sec.~\ref{sec:conc} the  conclusions of the work are drawn together along with the implications for future research.

\section{Formalism}
\label{sec:formalism}

\subsection{Modified Gravity Theory and Model}
\label{sec:MG}

A flat Friedmann-Roberston-Walker (FRW) metric in Newtonian gauge with sign convention $(-,+,+,+)$ is assumed
\begin{equation}
g_{\mu \nu} \mathrm{d}x^{\mu} \mathrm{d}x^{\nu}=a^2(\tau)\left[-(1+ 2 \Phi)\mathrm{d}\tau^2 + (1-2 \Psi) \gamma_{i j} \mathrm{d}x^i \mathrm{d}x^j\right]
\end{equation}
in which $\Phi$ is the Newtonian gravitational potential, $\Psi$ is the spatial curvature perturbation and $\gamma_{i j}$ is the 3D spatial metric. The spatial comoving coordinates are given by $x^{i}$ with $i,j$ running from 1 to 3. $\mu,\nu$ run from 0 to 3 including $\tau=x^0$, the conformal time defined by $d\tau=dt/a$, where $a(\tau)$ is the cosmological scale factor normalized to $a=1$ today.

In $f(R)$ gravity, the Einstein-Hilbert action \citep{Nojiri:2006gh} is replaced by
\begin{equation}
S=\int d^{4}x \sqrt{-g}\left(\frac{1}{16\pi G}[R+f(R)] + \mathcal{L}_{m}(\psi_{i})\right)
\label{eq:Action}
\end{equation}
where $f(R)$ is a function of the Ricci scalar, $R$. In this paper we consider the form of $f(R)$ proposed by Hu and Sawaki \citep{Hu:2007nk}, one of the most widely studied $f(R)$ models in the literature, in which the modification takes the form:
\begin{equation}
f(R) = -m^2 \frac{c_1 \left(R/m^2\right)^n}{c_2 \left(R/m^2\right)^n+1},
\label{eq:HuSfR} 
\end{equation}
with effective mass scale $m=H_0\sqrt{\Omega_{m0}}$, $H_0$ the Hubble constant, $\Omega_{m0}$ the fractional energy density in matter today and $c_1, c_2$ and $n$ free parameters in the model. 

By varying equation \eqref{eq:Action} with respect to the metric, one obtains the modified Einstein equations, 
\begin{equation}
G_{\mu \nu} + f_{R}R_{\mu \nu}-g_{\mu \nu}\left[\frac{1}{2}f(R)-\Box f_{R}\right]-\nabla_{\mu}\nabla_{\nu}f_{R} = 8 \pi G T_{\mu \nu}
\label{eq:Gmunu}
\end{equation}
where $\Box=g^{\mu \nu}\nabla_{\nu}\nabla_{\mu}$ and $f_R\equiv\frac{df(R)}{dR}$ is given in the high curvature regime limit, $R\gg m^2$, by 
\begin{equation}
f_R \simeq -n \frac{c_1}{c_2^2} \left(\frac{m^2}{R}\right)^{n+1}.
\label{eq:simplefR}
\end{equation}

Contracting (\ref{eq:Gmunu}) with $g^{\mu \nu}$ gives the trace equation, 
\begin{equation}
\Box f_{R}=\frac{1}{3}(R-f_{R}R+2f(R)-8\pi G \rho_{m}),
\label{eq:boxfR}
\end{equation}
where the subhorizon limit is assumed and ${T^\mu}_\mu=-\rho_m$ is taken to be dominated by cold dark matter. Equation (\ref{eq:boxfR}) can be viewed as an equation of motion for the scalar field $f_R$ with the right hand side acting as a driving term from an effective potential $\frac{dV_{\mathrm{eff}}}{df_R}$. 

Requiring that the background expansion history match that from $\Lambda$CDM further constrains the Hu-Sawicki model parameters. For a $\Lambda$CDM expansion history, one relates the background value of the Ricci scalar to the cosmological matter composition,
\begin{equation}
\bar{R}=3m^2\left(a^{-3}+4\frac{\Omega_{\Lambda 0}}{\Omega_{m0}}\right),
\label{eq:Rbar}
\end{equation}
where $\Omega_{\Lambda 0}$ is the energy density of a cosmological constant that would give rise to the observed expansion history. In tandem with minimizing $V_{\mathrm{eff}}$, $\frac{dV_{\mathrm{eff}}}{df_R}=0$, this fixes $\frac{c_1}{c_2}=6\frac{\Omega_{\Lambda}}{\Omega_{m}}$, leaving $\frac{c_1}{c_2^2}$ and $n$ as the remaining free model parameters. 

It is customary in the literature to not specify $\frac{c_1}{c_2^2}$, but instead specify $f_{R0}$, or the background field value at $z=0$,
\begin{equation}
\bar{f}_{R0} \simeq -\frac{n c_1}{c_{2}^{2}}\left[3\left(1+4\frac{\Omega_{\Lambda0}}{\Omega_{m0}}\right)\right]^{-(n+1)}.
\label{eq:fR0}
\end{equation}
  
In this analysis, $n=1$ and two values of $\vert f_{R0}\vert =10^{-6}$ and $10^{-5}$ are considered, referred to as F6 and F5, respectively.
 
Under these assumptions, and noting for the Hu-Sawicki model, $\bar{f_R} \ll 1$ and $\delta R \simeq \frac{dR}{df_R} \delta f_R \gg \bar{R} \delta f_R$, Eq. \eqref{eq:boxfR} can be  simplified, giving
\begin{equation}
\nabla^2 f_R = \frac{1}{3}a^2 \delta R(f_R)-\frac{8}{3}a^2 \pi G \bar{\rho}\delta,
\label{eq:dfR}
\end{equation}
where $\delta X$ denotes perturbations in a quantity X relative to the homogeneous background value, $\nabla^2 =\gamma^{i j} \nabla_i \nabla_j$ (after imposing the quasistatic approximation) and $\delta\equiv\delta\rho/\bar{\rho}$. The remaining perturbed Einstein equations lead to

\begin{equation}
  \nabla^2 \Phi = \frac{16}{3} \pi G a^2 \bar{\rho} \delta-\frac{1}{6} a^2 \delta R(f_R).
  \label{eq:totalPhi}
\end{equation}
Equations \eqref{eq:dfR} and \eqref{eq:totalPhi} together completely specify the total gravitational potential $\Phi$. To highlight the phenomenology at play, one can compare the modified gravity model to that of regular GR, by defining an effective Newtonian potential that would be derived using the standard Poisson equation in GR, in the subhorizon limit, 
\begin{equation}
\nabla^2 \Phi_N=4 \pi G a^2 \bar{\rho}\delta.
\label{eq:Newton}
\end{equation}

Test particles moving along modified geodesics of the metric of the Jordan frame will experience a total gravitational force per unit mass given by 
\begin{equation}
\vec{g}_{total} \equiv - \frac{1}{a^2}\vec{\nabla} \Phi=-\frac{1}{a^2}\vec{\nabla} \Phi_N+\frac{1}{2a^2}\vec{\nabla} f_R \equiv \vec{g}_{N} + \vec{g}_{5},
\label{eq:Total}
\end{equation}
where $\vec{\nabla}$ again is the (spatial) comoving gradient arising from $\gamma_{ij}$. On its surface, \eqref{eq:Total} may look as though $f_R$ is acting to decrease the gravitational force from its Newtonian value, however this is not the case. Looking at \eqref{eq:dfR} and \eqref{eq:Newton}, one can see that $f_R$ and $\Phi_N$ have couplings to matter of the opposite sign, meaning that in the presence of a spherical overdensity, $-\frac{1}{a^2}\vec{\nabla} \Phi_N$ and $+\frac{1}{2a^2}\vec{\nabla} f_R$ will both point towards the matter source, so that gravity is enhanced relative to its Newtonian value. Another way to see this is to rewrite \eqref{eq:totalPhi} as 
\begin{equation}
\nabla^2 \Phi = \frac{4}{3} \nabla^2 \Phi_N-\frac{1}{6} a^2 \delta R(f_R).
\label{eq:4/3eqn}
\end{equation}

Physically, $\delta R(f_{R})$ acts as an environment-dependent mass term in the field equation for $f_{R}$ \cite{Hern_ndez_Aguayo_2018}. In this form, it is clear that gravity is at most enhanced by $1/3$ from its Newtonian value, with $\delta R(f_R)$ acting to decrease that enhancement.

Using  \eqref{eq:simplefR}, and writing $f_{R}=\bar{f}_{R} + \delta f_{R}$ explicitly,
\begin{equation}
\delta R=\left(\frac{\bar{f}_{R0}}{\bar{f}_{R}+ \delta f_{R}}\right)^{\frac{1}{n+1}}\bar{R}_0-\bar{R},
\label{eq:deltaR}
\end{equation}
which is nonlinear in $\delta f_R$. These nonlinearities are responsible for the ``chameleon'' mechanism \citep{Khoury:2003aq,Khoury:2003rn}, which greatly suppresses the fifth force in high density environments. 

A positive $\delta$ entering as a source into  \eqref{eq:dfR} will act to make $\delta f_{R}$ positive due to the negative matter coupling. Given (\ref{eq:deltaR}), $f_{R}$ must be strictly negative, so that overdense regions with $\delta>0$ push $\delta f_{R}$ positive which causes the combination $\bar{f}_{R} + \delta f_{R}$ to grow smaller in magnitude, thereby turning on the nonlinearities contained in $\delta R$. We note that, depending on the model's value of $\bar{f}_{R0}$, high density may not necessarily imply high curvature as shown in \cite{PhysRevD.90.103505}, indicating that the degree of screening is highly dependent on the specific value of $\bar{f}_{R0}$ for a given model.

Taking into account the sign requirement, 
 \begin{equation}
 f_{R}= -\vert\bar{f}_{R0}\vert \left(\frac{\Omega_{m,0} + 4 \Omega_{\Lambda,0}}{a^{-3}\Omega_{m,0} + 4 \Omega_{\Lambda,0}}\right)^{n+1} + \delta f_{R}.
 \label{eq:barPlusDelta}
 \end{equation}

The interior of void regions feature a negative $\delta \rho$ which pushes the $\delta f_{R}$ field to a negative value, thereby gradually turning off the screening mechanism and enhancing the modifications to gravity. Since the source term in  (\ref{eq:dfR}) pushes $\delta f_{R}$ negative, and thus away from the nonlinear effects, we can linearly approximate $\delta R \simeq \frac{\mathrm{d}R}{\mathrm{d}f_R}\delta f_R$, and  (\ref{eq:dfR}) becomes
\begin{equation}
 \nabla^2 f_{R,lin} = a^2 \mu^2 \delta f_{R,lin} - \frac{8}{3} \pi G a^2 \bar{\rho} \delta,
 \label{eq:dfRlin}
\end{equation}
 with the background scalar mass given as 
\begin{align}
\mu^2 & = \frac{1}{3(n+1)}\frac{\bar{R}}{\vert \bar{f}_{R0}\vert}\left(\frac{\bar{R}}{\bar{R}_0}\right)^{n+1} \nonumber \\
 & = \left(\frac{1}{2997}\right)^2 \frac{1}{2\vert \bar{f}_{R0}\vert }\frac{(\Omega_{m0}a^{-3}+4\Omega_{\Lambda 0})^{n+2}}{(\Omega_{m0} +4\Omega_{\Lambda 0})^{n+1}} \ [(h/Mpc)^{2}].
\end{align}

To quantitatively capture the difference between the full and linearized fifth forces, we introduce a ``screening factor'', $\alpha(\vec{x})$,   defined  through, 
\begin{equation}
\vec{\nabla} f_{R,full} = \alpha(\vec{x}) \vec{\nabla} f_{R,lin}.
\label{eq:alpha}
\end{equation}
The total gravitational force can then be written as
\begin{equation}
\vec{g}_{total}=-\frac{1}{a^2}\vec{\nabla} \Phi_N+ \alpha \frac{1}{2a^2}\vec{\nabla} f_{R,lin}.
\label{eq:TotalLin}
\end{equation}

Respectively, $f_{R,lin}$ and $\alpha$ each speak to different aspects of the physics contained in the full nonlinear field equation (\ref{eq:dfR}), and provide complementary perspectives on the modified gravity phenomenology in voids. 

\subsection{Cosmological Simulations}
\label{sec:sims}

In this paper we use the N-body ELEPHANT (Extended LEnsing PHysics using ANalaytic ray Tracing) simulations described in \citep{Cautun_2018,alam2020testing}. The ELEPHANT simulations were created using the code N-body code ECOSMOG \citep{Li_2012,Bose_2017}, which itself is based on the gravitational N-body code RAMSES \cite{Teyssier:2001cp}. The code uses an adaptive mesh, which is refined based on the local density of particles in order to numerically solve the nonlinear field equation \eqref{eq:dfR} accurately.

We consider 5 sets of initial conditions, each realized at $z_{i} = 49$, and evolved forward until $z=0$ using either $GR$ (baseline), $F6$ (weakly modified) or $F5$ (strongly modified) cosmologies. Each simulation has a volume of $1024^3 (Mpc/h)^3$ and features $1024^3$ dark matter particles of equal mass. 

The cosmological parameters are chosen to match those from the 9-year WMAP release \cite{Hinshaw_2013}, namely $\Omega_{b}=0.046$, $\Omega_{c}=0.235$, $\Omega_{m}=0.281$, $\Omega_{\Lambda}=0.719$, $h=0.697$, $n_{s}=0.971$, and $\sigma_{8}=0.820$. 

\subsection{Void Identification and Classifications}
\label{sec:voidident}

Voids are identified using the void finder VIDE (Void IDentification and Examination toolkit) \citep{Sutter:2014haa}. VIDE implements an enhanced version of the void finding algorithm ZOBOV (ZOnes Bordering On Voidness) \citep{Neyrinck:2007gy}.
ZOBOV is a parameter free void finding algorithm which uses Voronoi tessellation followed by a watershed algorithm to identify voids. Each void is assigned an effective radius,
\bea
R_{\mathrm{eff}}=\left(\frac{3V_{void}}{4\pi} \right) ^{1/3} ,
\eea
where $V_{void}$ is the comoving void volume according to the watershed transformation, which means that we also always take $R_{\mathrm{eff}}$ to be comoving. Each void is also assigned a ``macrocenter" (from hereon referred to as center), which is given by,
\begin{equation}
\vec{X}_{v}=\frac{1}{\sum_i V_i} \sum_i \vec{x}_i V_i
\label{eq:macrocenter}
\end{equation}
where $\vec{x}_{i}$ is the comoving position of the $i^{th}$ halo in the void and $V_{i}$ is the corresponding cell volume assigned to each halo during the Voronoi tessellation. The sum is taken over all halos whose Voronoi cells constitute the same void. All position and velocities in our analysis are in real space as opposed to redshift space.

Voids are located using halo data (identified using the Rockstar halo finding algorithm \citep{2013ApJ...762..109B}) rather than the underlying particle data, to most closely align with astrophysical observables.

The default VIDE criteria for a void is  any ``catchment basin" identified by the watershed transform with an average number density within $r=0.25R_{\mathrm{eff}}$ from the center is less than $0.2 \bar{n}$ as determined from the halo data. Analyses \cite{Nadathur_2015A, Nadathur_2015B, 10.1093/mnras/stv513} that utilize the VIDE prescription have shown that this criteria is too strict, and that it can make void identification highly susceptible to Poisson fluctuations, which can exclude well-defined void regions because of the presence of a single halo  within $0.25 R_{\mathrm{eff}}$. Following these authors, we do not impose the central density criteria, and consider all local ``catchment basins" as voids in our analysis. We have, however, checked that imposing the criteria does not alter the findings in our work, beyond  the smaller void sample increasing the signal covariance.

Subvoids, or ``child" voids as identified by VIDE, are not considered in this work in an attempt to keep the analysis focused on void environments which are as uniform as possible. 

\begin{table}[b!]
 	\begin{tabular}{ | C{5em} | C{3.0em} | C{3.0em} |C{3.0em}||C{3.0em} | C{3.0em} |C{3.0em}| }
		\hline 
		\multirow{3}{*}{$R_{\mathrm{eff}} 	(Mpc/h)$} & \multicolumn{3}{c||}{\% of all voids}& \multicolumn{3}{c|}{\% of voids in this $R_{\mathrm{eff}}$  } 
		\\
		 & \multicolumn{3}{c||}{in this $R_{\mathrm{eff}}$ bin}& \multicolumn{3}{c|}{bin that are \Rtype } 
		\\ \cline{2-7}
			& GR 	& F6		& F5		& GR 	& F6		& F5
		\\ \hline
		5 - 15 	& 18\%	& 20\% 	& 20\% 	& 47\% 	& 47\%	&	47\%	 
		\\ \hline
		15 - 25 	& 50\% 	& 51\% 	& 51\% 	& 53\%	& 52\%	& 	53\%
		 \\ \hline
		 25 - 35 	& 22\% 	& 20\% 	& 20\% 	&56\%	 & 57\%	&	57\%
		 \\ \hline		 
		 35 - 45 	& 6\% 	& 5\% 	&5\%	& 67\%	& 69\% 	&	71\%
		 \\ \hline		 
		 45 - 55 	& 2\% 	&2\% 	& 2\% 	&75\% 	& 79\% 	&	78\%
 		 \\ \hline\hline	
		 All voids	& \multicolumn{3}{c|}{}		& 54\%	& 54\%	&	55\%
		 \\\cline{1-1}\cline{5-7}
 	\end{tabular}
	
 	\caption{Summary of the properties of voids identified at $z=0.5$ across the five realizations of each of the three cosmological models, $GR$, $F6$ and $F5$. The distribution of voids for each cosmology as a function of size, parameterized by the effective radius $R_{\mathrm{eff}}$ is shown [left columns], as is the fraction of voids in each size bin that are identified as \Rtype [right columns].}
 	\label{tab:tab1}
 \end{table} 

Following Ceccarelli \textit{et al.}~\citep{Ceccarelli:2013rza}, and similar to other authors \citep{2015MNRAS.454..889N,2017MNRAS.467.4067N}, one can classify voids based on their density profiles. Heuristically, \Stype (S for ``shell") voids are those in which the central void region is surrounded by a large overdense shell, whereas \Rtype (R for ``rising") voids feature a much smaller shell in comparison, remain underdense for a larger $r/R_{\mathrm{eff}}$ range, and more smoothly rise to the background density. The \Rtype and \Stype characterizations are respectively aligned with the void-in-void and void-in-cloud descriptions proposed by Sheth and van de Weygaer \citep{Sheth2004AHO}. 

Each void is classified by considering the average integrated density, $\Delta$, obtained from the void radial density profile as defined by the halo distribution. $\Delta$ is defined as
\begin{equation}
\Delta(R) = \frac{1}{\frac{4}{3} \pi R^3} \int_0^{R} 4 \pi r^2 \bar{\delta}(r) \, \mathrm{d}r,
 \label{eq:Delta}
\end{equation}
where $r$ is the radial coordinate taken from each void center, $\bar{\delta}(r)$ is the average halo number density contrast of the shell at radius $r$, and $R$ is the integral cutoff, given in terms of $R_{\mathrm{eff}}$. To classify each void, we average out to a cutoff of $R = R_{\mathrm{eff}}$  and identify those voids with $\Delta(R_{\mathrm{eff}})<0$  as $R$-type, and those with $\Delta(R_{\mathrm{eff}})>0$ as $S$-type. The sensitivity of the analysis to the cutoff scale was assessed by varying it out to $1.3R_{\mathrm{eff}}$; none of the central results depend on the value in this range.

\begin{figure*}[!t]
\includegraphics[width=0.9\linewidth]{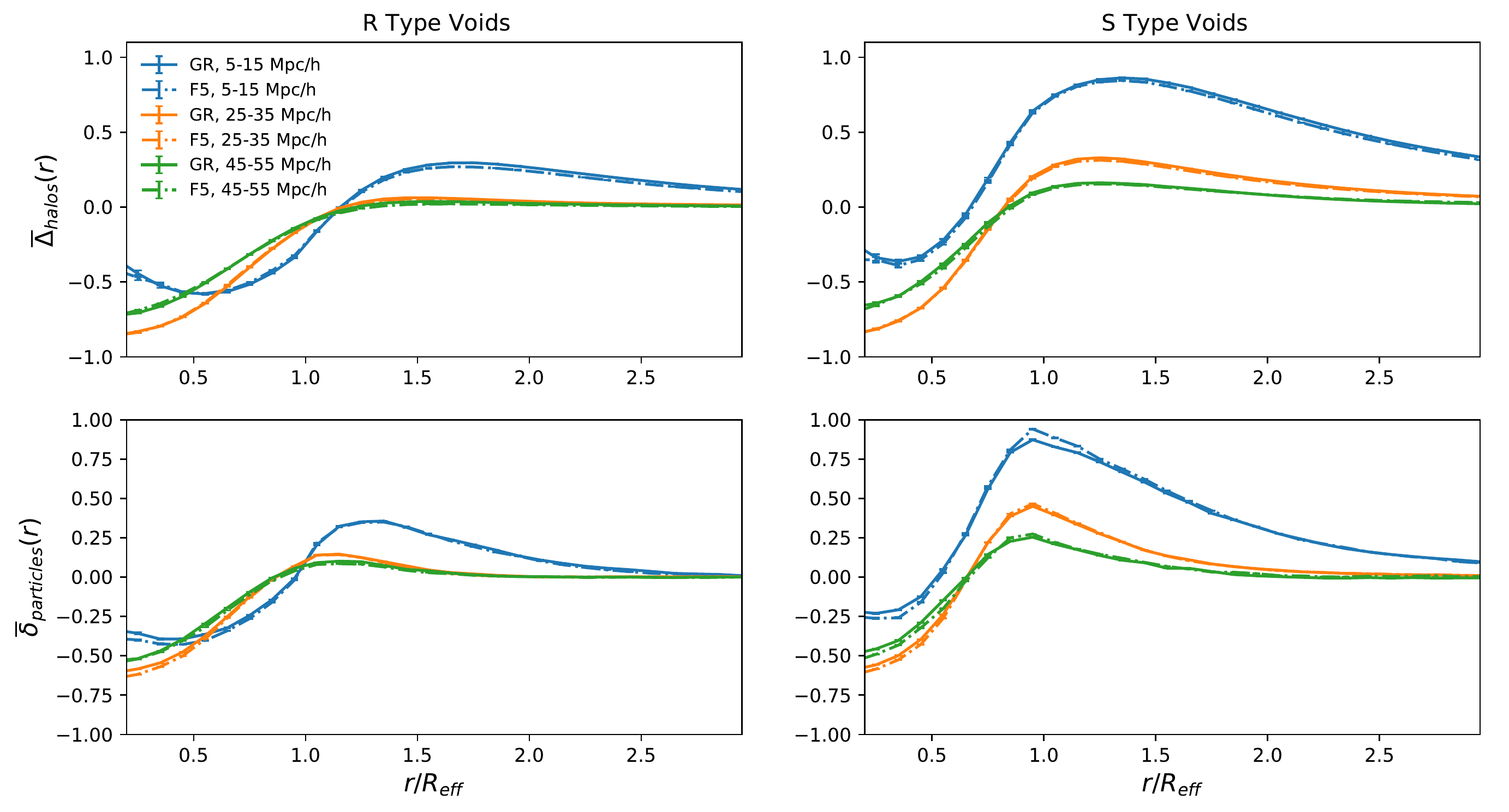}

\caption{The integrated number density contrast profiles from the halos $\bar\Delta_{halos}$ [upper panels] and the unintegrated number density contrast profiles from the particles $\bar\delta_{particles}$ [lower panels], averaged across all voids, for the \Rtype [left panels] and \Stype classifications [right panels] in $GR$ [full line] and $F5$ [dot dashed] at redshift $z=0$. Voids are binned by size: $R_{\mathrm{eff}}=5-15Mpc/h$ [blue], $25-35 Mpc/h$ [orange] and $45-55 Mpc/h$ [green]. }
\label{fig:densitiesAcrossScale}
\end{figure*}

\section{Results}
\label{sec:results}

\subsection{Void Density Profiles}
\label{sec:densityprofiles}

Structure growth is promoted in $f(R)$ theories, with greater numbers of halos and higher masses. This also leads to the number of voids being enhanced \cite{Padilla:2014hea,Cai:2014fma}.

In this analysis we consider voids with $5Mpc/h <R_{\mathrm{eff}}<55 Mpc/h$, constituting approximately $98\%$ of all voids across simulations and redshifts.  Across five realizations and using halos as the density tracer, at $z=0$,  37,514 voids are found in $GR$ simulations, 42,093 voids in $F6$, and 44,941 voids in $F5$. Similarly at $z=0.5$, 40,508 voids are identified in $GR$, 45,534 voids in $F6$, and 47,987 voids in $F5$.

Table \ref{tab:tab1} summarizes the properties of the voids at $z=0.5$ for the three cosmologies.  The fractional distribution of voids as a function of size  is consistent across the scenarios, with just slightly less than three quarters of identified voids having $R_{\mathrm{eff}}<25Mpc/h$ regardless of cosmology. The divisions between $R$ and \Stype classifications are similar across the three cosmologies, with $\sim$50\% of the voids identified as \Rtype averaging over all scales, and with the fraction of \Rtype voids ranging from $\sim$45\% to $\sim$80\% as one moves from the smallest to largest voids. The fractional distributions as a function of size and morphology do not significantly change between $z=0.5$ and $z=0$, again regardless of model. It should be noted that although the smallest size bin extends down to $5 Mpc/h$, only approximately $15\%$ of voids in the $5-15Mpc/h$ bin are themselves smaller than $10Mpc/h$, with a mean comoving size of roughly $12.3 Mpc/h$, a trend that holds across redshift and cosmology. Other size bins are more uniform in their distributions.

The density contrast profile of each void is calculated using the average halo or particle number density contrast $\delta(r)= (n(r)-\bar{n})/\bar{n}$ in spherical shells around the void's center. While the density profile of each void can be computed from either the particle or halo data, the void center and radius are always determined from the halos. In this way, the void identification is aligned with observational tracers, and  also provides a consistent center to compare  the radial density and velocity properties derived from the halo and particle data. Although unobservable, density profiles from the particle data are important as they allow  a consistency check on the halo data, and a mechanism to determine the full gravitational potential within voids. 

Void density profiles are presented in Fig.~\ref{fig:densitiesAcrossScale} using a rescaled radial coordinate $\tilde{r}=r/R_{\mathrm{eff}}$ averaging sums across the voids in the simulated samples to mitigate Poisson noise, as outlined in \cite{10.1093/mnras/stv513}. Integrated density contrast profiles, $\Delta(\tilde{r})$, are shown from the halo data (used in the classification of voids) whereas unintegrated density contrast profiles $\delta(\tilde{r})$ are shown from the particle data (used to later calculate underlying gravitational forces) for $R$ and \Stype voids in $GR$ and $F5$ at $z=0$. The density profiles are found to have a common form across void sizes when expressed in terms of this $\tilde{r}$ radial coordinate, consistent with \cite{Ricciardelli:2014vga} and \cite{Hamaus_2014}.  This common form is also shared with the $F6$ voids, which are not shown. \Rtype voids with $R_{\mathrm{eff}}>15Mpc/h$ have an average density profile which smoothly rises from an interior underdense region to an external region of essentially mean density. The smallest \Rtype voids feature some qualitative differences when compared to the larger \Rtype voids, with a smaller interior under density and an overdense shell at $r>R_{\mathrm{eff}}$. The \Stype voids consistently feature a large overdense shell, peaking at $r\sim R_{\mathrm{eff}}$, and  dwarfing that of their \Rtype counterparts. As one moves from smaller to larger voids, the density profiles of both $R$ and \Stype voids begin to have smaller overdense shells, consistent with the profiles shown in \cite{Hamaus_2014}.

The particle density data at $r<R_{\mathrm{eff}}$ are consistent with the findings in \cite{Zivick:2014uva} in which $f(R)$ gravity is found to have ``emptier, more steeply-walled voids". The halo profiles show less pronounced differences between $GR$ and the modified theories. The relative importance of the small differences between the particle density profiles and the fifth forces in $F5$ and $F6$ to the halo radial velocities will be considered later.

\subsection{Radial Velocity Profiles}
\label{sec:radvel}

\begin{figure*}[!t]
{\includegraphics[width=1\linewidth]{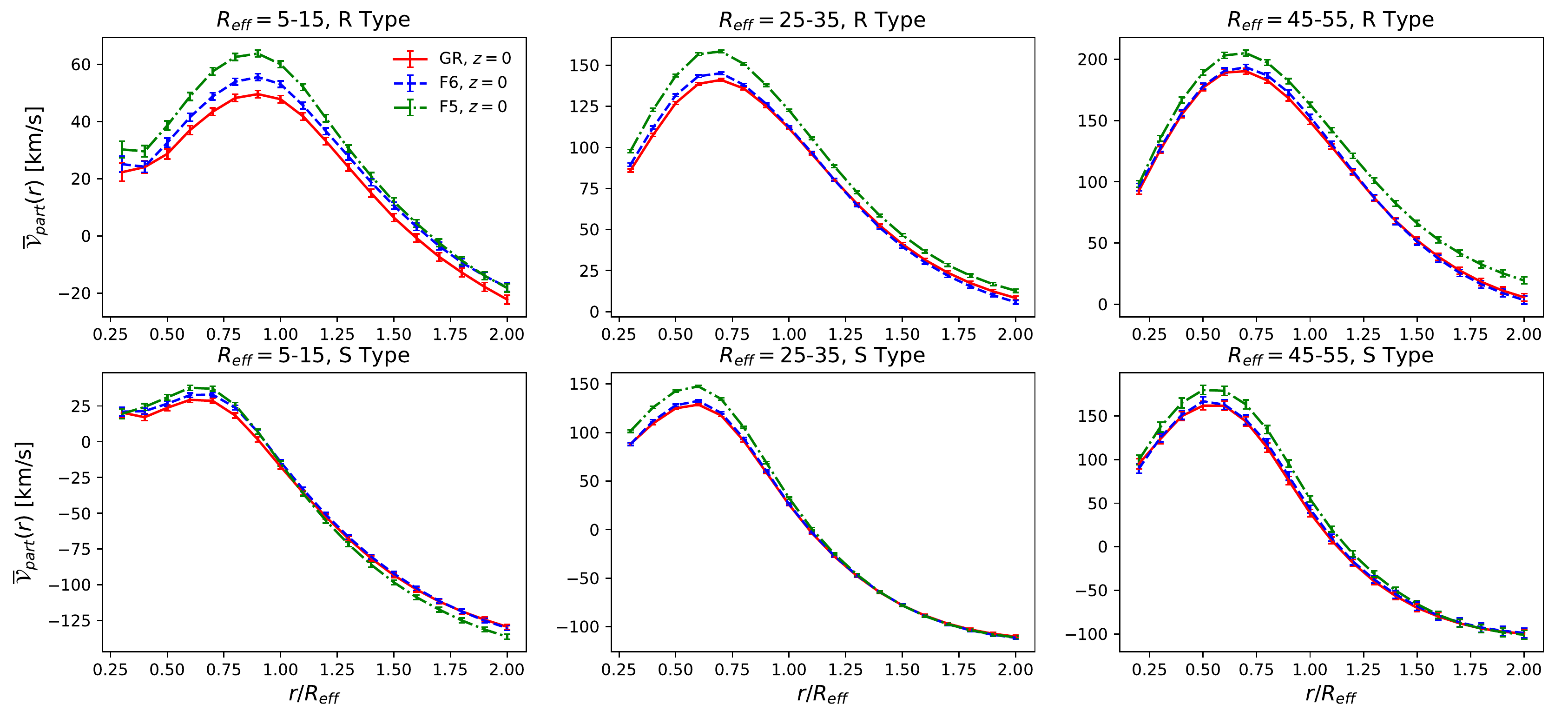}}
\caption{The mean integrated radial velocity profiles from the particle data, $\bar{\mathcal{V}}_{part}$, for \Rtype voids [upper panels] and \Stype voids[lower panels] at redshift $z=0$. Data for voids of size $R_{\mathrm{eff}}$=5-15$Mpc/h$ [left], 25-35$Mpc/h$ [center], and 45-55$Mpc/h$ [right] are shown for GR [full, red line], F6 [blue dashed line] and F5 [green dot-dashed line] .}
\label{fig:partRadVelz0p0}
\end{figure*}

Given the suggested challenges in differentiating between GR and modified gravity cosmologies with the halo density profiles alone, we now consider the potential of a second observable statistic, the void radial velocity profiles. As with the density profiles, the velocity profiles can be constructed from either the simulated halo or particle data separately. Doing so allows us to perform a consistency check between the biased tracers and the CDM particle distribution. For a given void, and given comoving distance $r$ from the void center, the radial velocity profile is computed by averaging over all tracers interior to $r$. This integrated measure maximizes the signal to noise relative to considering individual radial shells, especially when halos are considered. The integrated radial velocity profile is given by 
\begin{equation}
\bar{\mathcal{V}}(r)= \frac{1}{N(r)} \sum_{j} \theta(r-\vert \vec{x}_j-\vec{X}_{v}\vert ) \ \vec{v}_j\cdot \hat{r}, 
\end{equation}
where $\vec{x}_{j}$ and $\vec{v}_{j}$ are respectively the position and peculiar velocity of the $j^{th}$ tracer (halo or particle), $\vec{X}_{v}$ is the void center, $r$ is the comoving distance from $\vec{X}_{v}$ to the edge of spherical region being averaged over, $\hat{r}$ is the radial unit vector, $\theta$ is the Heavyside function and $N(r)=\sum_{j} \theta(r-\vert \vec{x}_j-\vec{X}_{v}\vert )$ is the total number of tracers interior to the radial coordinate $r$. We neglect halos within 2.5$Mpc/h$ of the void center from the analysis, as when taking the radial component of the velocity, the inner most halos are the most affected by potential uncertainties in the halo-determined void center.

Figure \ref{fig:partRadVelz0p0} gives the radial velocity profiles derived from particle data at $z=0$ for the three cosmologies, separating voids by size and classification. Within each void, the outflow velocities for both $S$ and \Rtypes increase in magnitude with increasing void size, however in larger voids the outflows peak at smaller $\tilde{r}=r/R_{\mathrm{eff}}$. The particle velocity profiles at $z=0$ across the three cosmologies are distinct in \Rtype voids at all sizes with the exception of $GR$ and $F6$ in the largest voids with the strength of the outflow correlated with the strength of the modification to gravity. We find that the differences are most pronounced in the smallest voids, $R_{\mathrm{eff}}<15Mpc/h$, with both F6 and F5 models distinguishable from GR at the peak of the outflow at $r \sim 0.9 R_{\mathrm{eff}}$. While the intermediate scale voids with $R_{\mathrm{eff}}\approx 20-40 Mpc/h$ are most numerous, the relative differences in the outflows are much smaller than that of the smallest voids. While we find differences between GR and F5 in the largest voids, we are unable to distinguish between GR and F6 in the largest \Rtype voids, $R_{\mathrm{eff}}>40 Mpc/h$, however these voids are far less numerous, as shown in Table \ref{tab:tab1}, and therefore the sample variance is greater.

By comparison, velocity profiles in the small and intermediate \Stype voids, with $R_{\mathrm{eff}}<40 Mpc/h$, do not show significant differences across the three cosmologies, especially between $GR$ and $F6$. For a given void size, the outflows are of the same magnitude across cosmologies and are limited to the void interiors, $r\lesssim R_{\mathrm{eff}}$. We do find velocity profile differences between GR and F5 in the large \Stype voids concentrated well inside the void, at $r\lesssim 0.75 R_{\mathrm{eff}}$, but again are unable to use these voids to distinguish between GR and F6.

\begin{figure*}[!t]
\includegraphics[width=1\linewidth]{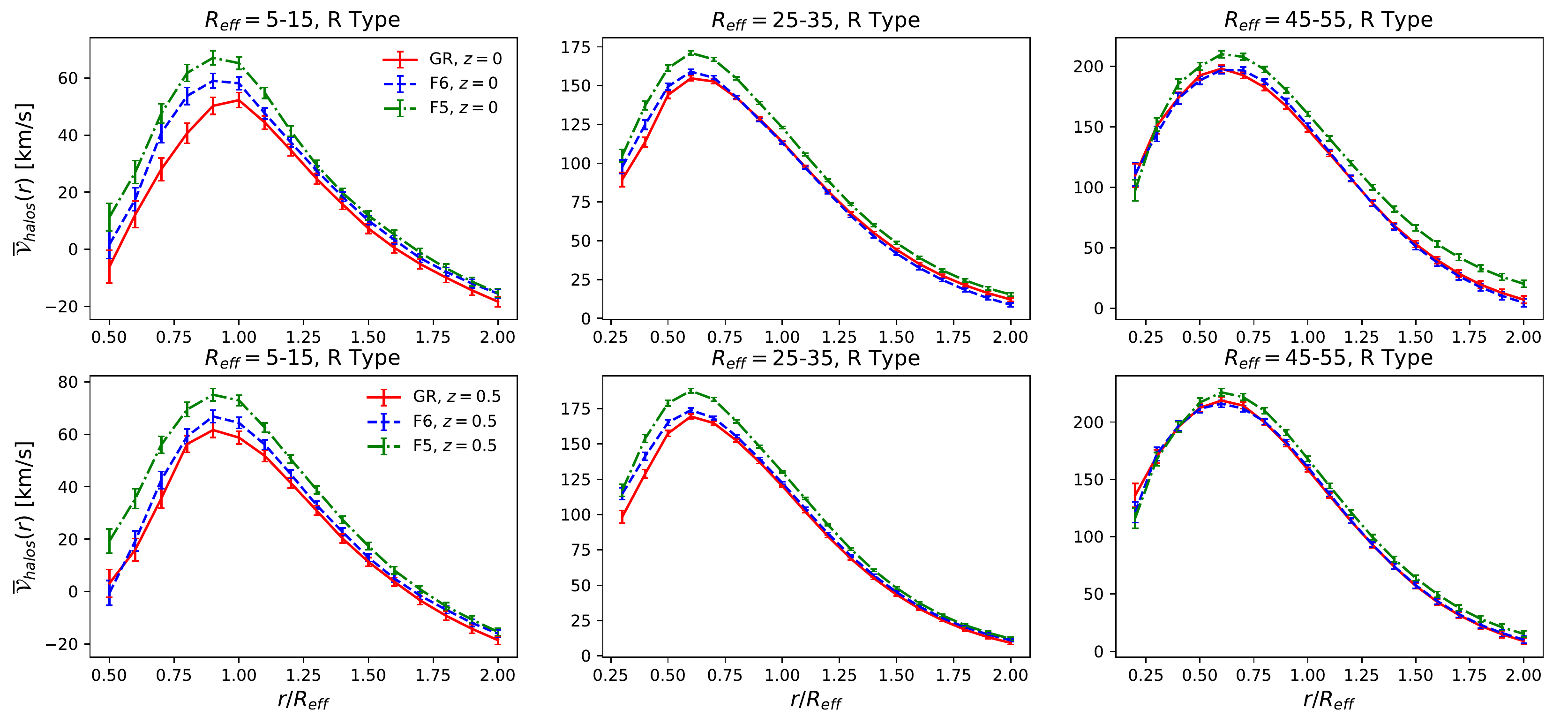}

\caption{The mean integrated radial velocity profiles from the halos, $\bar{\mathcal{V}}_{halos}$, for \Rtype voids at redshift $z=0$ [upper panels] and $z=0.5$ [lower panels]. Data for voids of size $R_{\mathrm{eff}}=5-15Mpc/h$ [left], $25-35Mpc/h$ [center], and $45-55Mpc/h$ are shown for GR [full, red line], F6 [blue dashed line] and F5 [green dot-dashed line].}
\label{fig:haloRadVelBoth}
\end{figure*}

Figure \ref{fig:haloRadVelBoth} shows the integrated velocity profiles derived from halos in \Rtype voids at $z=0$ and $z=0.5$. We find trends consistent with those shown in the particle data -- the outflow component of the velocity profiles in small \Rtype voids again offer the best opportunity to differentiate between the three cosmologies while the distinguishing power of the other larger sized voids falls off with increasing $R_{\mathrm{eff}}$.

\begin{figure*}[!t]
{\includegraphics[width=1\linewidth]{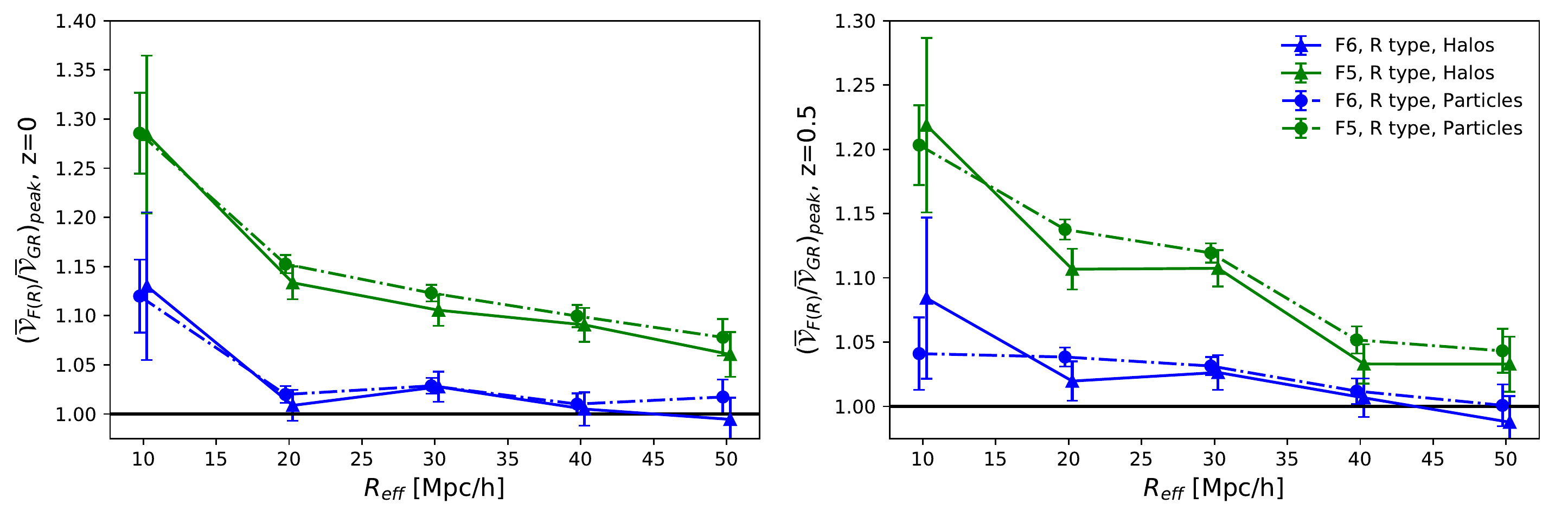}}
\caption{The ratio of the mean peak radial velocities, $\bar{\mathcal{V}}$, in the \Rtype voids  for the F5 [green lines] and F6 [blue lines] cosmologies relative to those observed in GR are compared at $z=0$ [left panel] and $z=0.5$ [right panel]. Results derived from the halos [full lines] and particles [dot-dashed lines] are given as a function of void size, $R_{\mathrm{eff}}$.  }
\label{fig:velRatioPlots}
\end{figure*}

The effects of modified gravity on the \Rtype velocity profiles as a function of void size are summarized in Fig.~\ref{fig:velRatioPlots} by showing the ratio of the mean integrated velocity in the modified gravity models relative to that in the GR, evaluated at the peak value of both. The peak velocity is modified most in the smallest 5-15 $Mpc/h$ voids. At $z=0.5$ the velocity ratio computed from the halos in these voids is $1.08 \pm 0.06$ for $F6$ and $1.22\pm 0.07$ for $F5$. The halo ratios are shown to be consistent with those results derived directly from the particles. As void size increases, the ratio of the peak values in $GR$ and $F6$ becomes more consistent with unity. 

\section{Analysis}
\label{sec:analysis}

To get a better intuition for the fifth force acting in void environments, the full fifth force can be understood in terms of the linearized fifth force from (\ref{eq:dfRlin}) and the screening (or enhancement) factor $\alpha$ using \eqref{eq:alpha}. 
Analysis of the linearized field equation for various sizes and types of voids will inform us to how the fifth force contained in (\ref{eq:dfRlin}) interacts with different void scales and density profile shapes, while analysis of the screening factor, $\alpha$, will inform us of the effects of the nonlinear chameleon mechanism, and deviations from the forces obtained in the linearized limit.

\subsection{Interpretation Using the Linearized Fifth Force Equation}
\label{sec:linfifth}

Voids of the same classification display similar density profiles in terms of $\tilde{r}=r/R_{\mathrm{eff}}$, as shown in  Fig.~\ref{fig:densitiesAcrossScale}. Thus, to understand differences in the linearized fifth force within voids of the same classification, it is instructive to use the same normalized coordinate and its dimensionless reciprocal space equivalent, $\tilde{k}=k R_{\mathrm{eff}}$, giving
\bea
   \Phi_N(\tilde{k}) &=& -\frac{4 \pi G a^2 \bar{\rho} R_{\mathrm{eff}}^2}{\tilde{k}^2} \delta(\tilde{k}),
   \\
  \delta f_{R,lin}(\tilde{k})&=& \frac{\frac{8}{3} \pi G a^2 \bar{\rho} R_{\mathrm{eff}}^2}{\tilde{k}^2+ \mu^2 a^2 R_{\mathrm{eff}}^2} \delta(\tilde{k}).
\eea

Assuming spherical symmetry, this change of variables allows the linearized fifth force $\vec{g}_{5,lin}$ and the Newtonian force $\vec{g}_{N}$ in \eqref{eq:TotalLin} to be expressed as,
\bea
\vec{g}_N(\tilde{r}) \cdot \hat{r} &=&-\frac{1}{aR_{\mathrm{eff}}}\partial_{\tilde{r}} \left( \Phi_{N} \right) \nonumber
\\
&= &16\pi^2 G \bar{\rho} \int \tilde{k}^2\mathrm{d}\tilde{k} \delta(\tilde{k}) W_{N}(\tilde{k},R_{\mathrm{eff}},a) \partial_{\tilde{r}}\left(\frac{sin(\tilde{k}\tilde{r})}{\tilde{k}\tilde{r}}\right), \nonumber
\\
&&
\\
\vec{g}_{5,lin}(\tilde{r}) \cdot \hat{r} &=& \frac{1}{2aR_{\mathrm{eff}}}\partial_{\tilde{r}}\left(\delta f_{R,lin} \right) \nonumber
\\
 &=&\frac{16\pi^2 G \bar{\rho} }{3} \int \tilde{k}^2\mathrm{d}\tilde{k} \delta(\tilde{k}) W_{5}(\tilde{k},R_{\mathrm{eff}},a) \partial_{\tilde{r}}\left(\frac{\sin(\tilde{k}\tilde{r})}{\tilde{k}\tilde{r}}\right). \nonumber
 \\
&&
\eea
Here $\vec{g}(\tilde{r}) \cdot \hat{r}$ is the physical magnitude of the Newtonian or fifth force in the radial direction with $\hat{r}$ the physical radial \textit{unit} vector, not the comoving radial basis vector, which accounts for the factor of $1/a$ instead of $1/a^2$. The effects of void scale are encapsulated within what we will henceforth refer to as the window functions for the Newtonian force and fifth force, respectively: 
\bea
W_{N}(\tilde{k},R_{\mathrm{eff}})&=&\frac{ aR_{\mathrm{eff}} }{\tilde{k}^2},
\label{eq:WindowDefN}
\\
W_{5}(\tilde{k},R_{\mathrm{eff}},a)&=&\frac{aR_{\mathrm{eff}} }{\tilde{k}^2+a^2\mu^2(a) R_{\mathrm{eff}}^2}.
\label{eq:WindowDefF}
\eea

A heuristic understanding of the effect of scale can be obtained from by considering the window functions of the above integrals evaluated at the particular wave-mode $\tilde{k}_{0}$ around which $\tilde{k}^2_{0}\delta(\tilde{k}_{0})$ is peaked. Given the commonality of $\delta(\tilde{r})$, $\tilde{k}_{0}$ is not expected to  significantly change as one moves across different $R_{\mathrm{eff}}$ size bins for voids of a given classification. Thus, by considering how the window function $W_{5}$ evaluated at $\tilde{k}_{0}$ varies as a function of $R_{\mathrm{eff}}$ and redshift $z$, we can get a good idea for how the linearized fifth force varies with scale and redshift within each class of voids. 

\begin{figure*}[!t]
\includegraphics[width=1\linewidth]{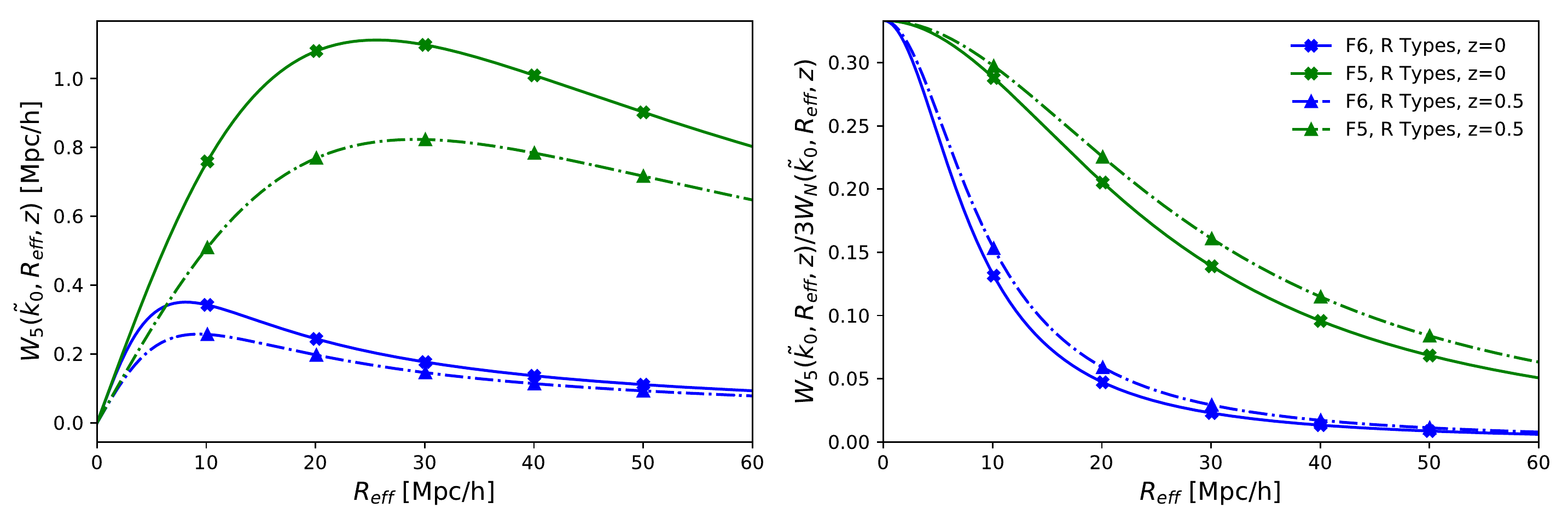}
\caption{[Left] The magnitude of the window function $W_{5}$ in \eqref{eq:WindowDefF} evaluated at the wave number $\tilde{k}_0$ where the reciprocal space quantity $\tilde{k}^2 \delta(\tilde{k})$ is peaked, for \Rtype voids at redshifts of $z=0$ [full line] and $z=0.5$ [dot-dashed line] for F5 [green] and F6 [blue]. [Right] The ratio of the fifth force and Newtonian window functions, in \eqref{eq:windowFunctionRatio}, evaluated at the same $\tilde{k}_0$.}
\label{fig:WindowFunction}
\end{figure*}

Fixing $\tilde{k}_{0}$, the window function for the fifth force is peaked at $R_{\mathrm{eff}}^{max} = \tilde{k}_{0}/a\mu$. At redshift $z=0.5$, we find for \Rtype voids in $F6$ that $\tilde{k}_{0}=3.5$, which when combined with the corresponding value of $\mu=0.56 h/Mpc$, translates to a peak in the window function at $R_{\mathrm{eff}}^{max}=9.3 Mpc/h$. For \Rtype voids in $F5$, we have a very similar value of $\tilde{k}_{0}=3.4$, but due to the smaller mass term of $\mu=0.18 h/Mpc$ we have a peak in the $F5$  window function at a larger value of $R_{\mathrm{eff}}^{max}=29Mpc/h$. This means that over the range of $R_{\mathrm{eff}}$ studied, the F6 window function decreases with increasing $R_{\mathrm{eff}}$, while in F5, the window function maintains a more consistent value. These properties are minimally affected by the change in redshift between $z=0$ and $z=0.5$. 

When thinking about potential observations, it can be useful to consider the relative, as well as the absolute, strength of the fifth force in comparison to the Newtonian gravitational force, to better assess the statistical distinguishability of the modified gravity effects. Given the similarities between the radial density profiles of voids of different sizes, the difference between the Newtonian gravitational force and the linearized fifth force can be effectively captured by the differences between the respective window functions,
\begin{equation}
\frac{W_{5}(\tilde{k}_0,R_{\mathrm{eff}},a)}{3 W_{N}(\tilde{k}_0,R_{\mathrm{eff}},a)}=\frac{\tilde{k}_0^2}{3(\tilde{k}_0^2+a^2 \mu^2(a)R_{\mathrm{eff}}^2)}
\label{eq:windowFunctionRatio}
\end{equation}
where the factor of $3$ is due to the relative factor of $3$ between the coupling constants. In the right-hand panel of Fig.~\ref{fig:WindowFunction} it is shown that, for a given value of $\tilde{k}_0$, the ratio is a strictly decreasing function of $R_{\mathrm{eff}}$. Physically, this is because, as $R_{\mathrm{eff}}$ changes, there is always tension between the Yukawa suppression acting to decreasing the fifth force (denominator of $W_{5}$), and the amount of under density sourcing the fifth field and increasing the fifth force (numerator of $W_{5}$), both of which increase with $R_{\mathrm{eff}}$. Since the Newtonian potential doe not suffer from Yukawa suppression (no mass term in denominator), $W_{N}$ strictly grows with $R_{\mathrm{eff}}$, and always at a faster rate than $W_{5}$. 

\subsection{Towards Solving the Full Nonlinear Fifth-Force Equation}
\label{sec:linfifthGreens}

In the discussion so far, the effects of scale have been highlighted by focusing on the peak of the density function in reciprocal space. Although this approach places the dependence on $R_{\mathrm{eff}}$ front and center, it neglects the contributions from the full $\delta(\tilde{r})$ profile, obscures the fact that the shell theorem has been explicitly violated, and does not extend to solving the full field in \eqref{eq:dfR}. Motivated to understand explicit effects of void shape, and to eventually solve \eqref{eq:dfR} exactly, \eqref{eq:dfRlin} is solved again but this time using a Green's function approach. 

Under the assumption of spherical symmetry, $\phi_{5}(\tilde{r},{\tilde{R}})$, the Green's function to the linearized field equation for a spherical matter shell located at $\tilde{R}$, is defined implicitly through 
\begin{equation}
\nabla_{\tilde{r}}^2 \phiG= a^2 \mu^2 R_{\mathrm{eff}}^2 \phiG - \frac{8}{3} \pi G a^2 \bar{\rho} R_{\mathrm{eff}}^2 \delta^{(1)}(\tilde{r}-\tilde{R})
\label{eq:Green'sFunctionDef}
\end{equation}
where $\delta^{(1)}(\tilde{r}-\tilde{R})$ is the $1D$ Dirac delta function, not to be confused with the density contrast $\delta(\tilde{r})$. Once  $\phiG$ is known, $\delta f_{R,lin}$ can be reconstructed via
\begin{equation}
\delta(\tilde{r})=\int \mathrm{d}\tilde{R} \left(\delta(\tilde{R}) \delta^{(1)}(\tilde{r}-\tilde{R})\right) \rightarrow \delta f_{R,lin} =\int \mathrm{d}\tilde{R} \left(\delta(\tilde{R})\phiG \right)
\label{eq:greenDef}
\end{equation}
Equation \eqref{eq:Green'sFunctionDef} can be solved analytically by standard methods involving contour integration, yielding 
\begin{equation}
\phiG=\frac{8 \pi a G \bar{\rho}}{3 \mu} \frac{\tilde{R}}{\tilde{r}} R_{\mathrm{eff}}
\begin{cases} 
  e^{-a \tilde{R}R_{\mathrm{eff}} \mu} \mathrm{sinh}(a \tilde{r}R_{\mathrm{eff}}\mu)
 & \text{, $\tilde{r} \leq \tilde{R}$ } 
 \\   
    {e^{-a \tilde{r}R_{\mathrm{eff}} \mu} \mathrm{sinh}(a \tilde{R}R_{\mathrm{eff}}\mu)} & \text{, $\tilde{r}>\tilde{R}$} \\
  \end{cases}
\label{eq:phiGdef}
\end{equation}

Since $\vec{g}_5(\tilde{r})=\frac{1}{2aR_{\mathrm{eff}}} \partial_{\tilde{r}} \left( \delta f_{R}\right) \hat r$, a new Green's function can be defined explicitly for the fifth force rather than for the field,
\bea
\mathcal{F}_{5}(\tilde{r},\tilde{R})&=&\frac{1}{2 a R_{\mathrm{eff}}} \partial_{\tilde{r}} (\phiG)
\nonumber
\\
&=& \frac{4 \pi G \bar{\rho}}{3 \mu} \tilde{R} \times\nonumber
\\
&&\hspace{-1.cm}\left\{
\begin{array}{ll}
  \frac{e^{-a \tilde{R} R_{\mathrm{eff}} \mu}}{\tilde{r}}\left[ (a R_{\mathrm{eff}} \mu ) \cosh(a \tilde{r} R_{\mathrm{eff}} \mu) - \frac{\mathrm{sinh}(a \tilde{r} R_{\mathrm{eff}} \mu)}{\tilde{r}} \right]
 & \text{, $\tilde{r} \leq \tilde{R}$ } 
 \\
  \frac{\sinh(a \tilde{R} R_{\mathrm{eff}} \mu)}{\tilde{r}} \left[-(a R_{\mathrm{eff}} \mu ) e^{-a \tilde{r} R_{\mathrm{eff}} \mu} - \frac{e^{-a \tilde{r} R_{\mathrm{eff}} \mu}}{\tilde{r}} \right]
 &\text{, $\tilde{r}>\tilde{R.}$} 
\end{array}\right. \nonumber
\\ 
  \label{eq:fifthFG}
\eea
For comparison the equivalent Green's functions for the Newtonian potential and force with spherical symmetry are the familiar functions 
\begin{equation}
\phi_{N}(\tilde{r},\tilde{R})= -4 \pi G \bar{\rho} a^2 R^2_{\mathrm{eff}} \tilde{R}^2   
\begin{cases} 
  \frac{1}{\tilde{R}}
 & \text{, $\tilde{r} \leq \tilde{R}$ } \\
 \frac{1}{\tilde{r}}
  & \text{, $\tilde{r}>\tilde{R}$} \\
  \end{cases}
\end{equation}
and using $\mathcal{F}_{N}(\tilde{r},\tilde{R})=-\frac{1}{aR_{\mathrm{eff}}} (\partial_{\tilde{r}} \phi_{N}(\tilde{r},\tilde{R}))$, 
\begin{equation}
\mathcal{F}_{N}(\tilde{r},\tilde{R})= -4 \pi G a \bar{\rho} R_{\mathrm{eff}} \tilde{R}^2   
\begin{cases} 
  0
 & \text{, $\tilde{r} \leq \tilde{R}$ } \\
 \frac{1 }{\tilde{r}^2}
  & \text{, $\tilde{r}>\tilde{R}$}\\
  \end{cases}
   \label{eq:newtonFG}
\end{equation}
recovering the shell theorem from Newtonian gravity. Comparing \eqref{eq:newtonFG} to \eqref{eq:fifthFG} for a given matter shell, the fifth force causes the attraction of a point particle both interior and exterior to the shell, whereas the Newtonian gravitational force only attracts an exterior particle. 

Focusing on $\tilde{r}<\tilde{R}$ in \eqref{eq:fifthFG}, the piece which explicitly violates the shell theorem,  for given values of $a$, $R_{\mathrm{eff}}$, and $\mu$ the average force interior to a mass shell will be maximized if the mass shell is placed at $\tilde{R}_{max}= \frac{2.73}{a R_{\mathrm{eff}} \mu}$. Considering this force has no Newtonian analog, maximizing this contribution to the fifth force will greatly enhance the ratio of $\vec{g}_{5}$ to $\vec{g}_{N}$. Looking at the void density profiles shown in Fig.~\ref{fig:densitiesAcrossScale}, we can see that our actual void density profiles have ``mass shells" of various sizes located at approximately $\tilde{R} \sim 1$. Plugging in values for $F6$ at $z=0.5$, we see that mass shells in this range are most effective if $R_{\mathrm{eff}}$ is taken to be $\sim 7 Mpc/h$ -- in reasonable agreement with the $R_{\mathrm{eff}}=9 Mpc/h$ estimate previously from the window function arguments. Repeating this calculation for $F5$ again at $z=0.5$, we find the $R_{\mathrm{eff}}$ which makes these shells most effective is  $\sim 23 Mpc/h$, relative to the earlier window function estimate of $R_{\mathrm{eff}}=29 Mpc/h$.

\subsection{Interpretation using the Nonlinear Fifth Force Equation}
\label{sec:nonlinfifth}

In the previous section we considered solutions to the linearized field equation. In order to understand the full response to the modified gravity theory we need to also determine whether the nonlinear solution differs significantly from the linearized one, as parameterized through the screening factor, $\alpha$. In this section we outline the iterative procedure we develop, using Green's functions, to solving the nonlinear field \eqref{eq:dfR} in voids.

For most voids it is expected that the nonlinear screening from the chameleon mechanism will be minimal in rare environments, and the linearized solution will be close to the full nonlinear solution. Thus, to begin the algorithm, the linearized field equation \eqref{eq:dfRlin} is solved using the Green's function method given by  \eqref{eq:greenDef} and \eqref{eq:phiGdef}  to obtain an initial estimate of the  full solution $f_{R,(0)} = f_{R,lin}$. In rare instances  in which the linear solution is unphysical, i.e., $\delta f_{R,lin}> (-\bar{f}_{R})$ so that $f_{R,lin}>0$, we smoothly modify the initial $\delta f_{R,(0)}$  such that it remains strictly negative, and sufficiently close to $(-\bar{f}_{R})$ to be an effective initial trial. The algorithm proceeds by modifying the current estimate at each iterative step until it is determined that it has converged to the full nonlinear solution. To characterize the degree to which the current estimate differs from the full solution, the $i^{th}$ iterative solution $f_{R,(i)}$ is plugged back into 
\eqref{eq:dfR}, and terms are rearranged in order to define a new density profile,
\begin{equation}
\delta_{(i)}=\frac{\nabla^2 f_{R,(i)} -\frac{1}{3}a^2 \delta R(f_{R,(i)})}{-\frac{8}{3}a^2 \pi G \bar{\rho}}.
\label{eq:1}
\end{equation}
In lieu of comparing $f_{R,(i)}$ to the full solution $f_{R}$, the latter of which is unknown, one can instead compare $\delta_{(i)}$ to the density function, $\delta_{real}$ from the particles in the simulation by defining
\begin{equation}
\epsilon_{(i)}(\tilde{r}) = \delta_{real}(\tilde{r})-\delta_{(i)}(\tilde{r}).
\label{eq:epsilon}
\end{equation}

If the difference between the iterative density estimate and that from the particles is greater than a desired tolerance, then we define a new field $\varphi_{(i)}$ defined as
\begin{equation}
\varphi_{(i)}=f_{R,full}-f_{R,(i)}.
\end{equation}
 Taking $\nabla^2_{\tilde{r}} \varphi_{(i)}$,
 \begin{equation}
 \nabla^2_{\tilde{r}} \varphi_{(i)} = \frac{1}{3}a^2 R_{\mathrm{eff}}^2 \left(\delta R(f_{R,(i)} + \varphi_{(i)}) - \delta R(f_{R,(i)}) \right)-\frac{8}{3}a^2 R_{\mathrm{eff}}^2 \pi G \bar{\rho}\epsilon_{(i)}
 \label{eq:phi}
 \end{equation}
Here, the linearization is done not around the background value of the field, but around $f_{R,(i)}$, such that \eqref{eq:phi} becomes
\begin{align}
\nabla^2_{\tilde{r}} \varphi_{lin,(i)} & = \frac{1}{3}a^2 R_{\mathrm{eff}}^2 \left(\frac{\delta R}{\delta f_{R}} \biggr \vert_{f_{R,(i)}}\right)\varphi_{lin,(i)} -\frac{8}{3}a^2 R_{\mathrm{eff}}^2 \pi G \bar{\rho}\epsilon_{(i)} \nonumber 
\\
& = a^2 R_{\mathrm{eff}}^2 \mu_{\mathrm{eff}}^2(\tilde{r}) \varphi_{lin,(i)}(\tilde{r})-\frac{8}{3}a^2 R_{\mathrm{eff}}^2 \pi G \bar{\rho}\epsilon_{(i)}(\tilde{r}),
\label{eq:phiLin}
\end{align}
with $\mu_{\mathrm{eff}}^2(\tilde{r})$ explicitly given as
\begin{equation}
\mu_{\mathrm{eff}}^2(\tilde{r})=\left(\frac{1}{3}\frac{\delta R}{\delta f_{R}} \biggr \vert_{f_{R}}\right)= - \frac{\bar{R}_{0}}{3(n+1)f_{R}(\tilde{r})}\left(\frac{\bar{f}_{R,0}}{f_{R}(\tilde{r})} \right)^{\frac{1}{n+1}}.
\label{eq:muEff}
\end{equation}

\begin{figure*}[!t]
\includegraphics[width=1\linewidth]{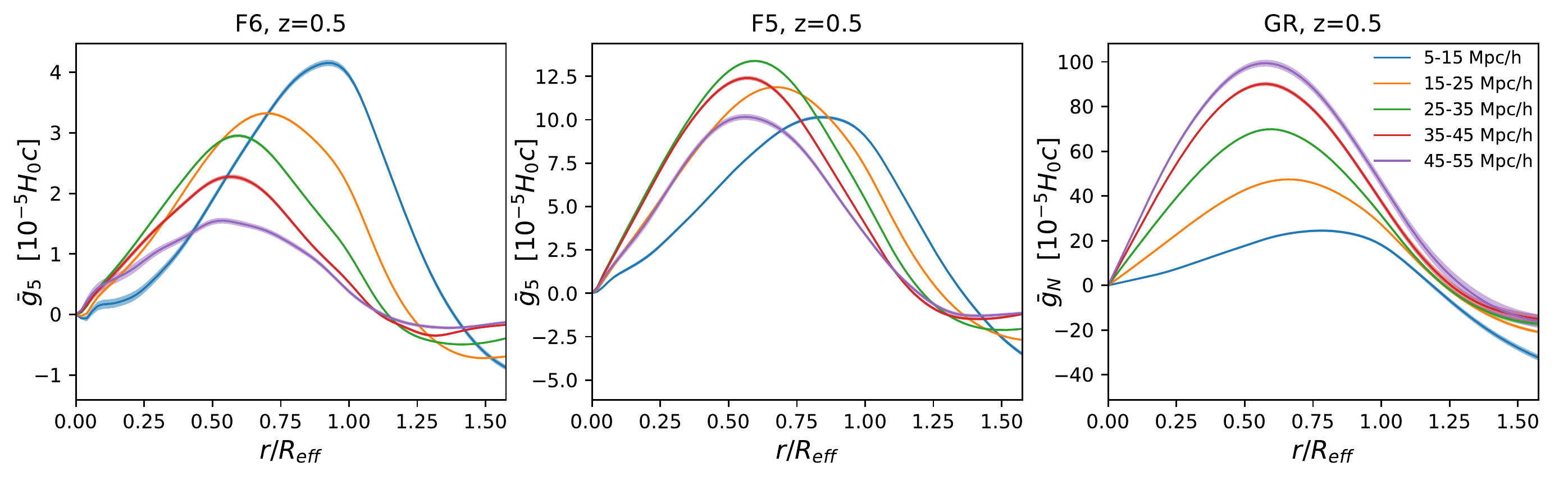}
\caption{The mean fifth force per unit mass, ${g}_5$, in \Rtype voids in $F6$ [left] and $F5$ [center] along with the Newtonian force, $g_N$, at $z=0.5$ along with the standard error on the mean [shaded regions]. The profiles are shown for voids increasing in size from 5-15$Mpc/h$ [blue lines] to 45-55$Mpc/h$ [purple lines]. }
\label{fig:fifthForces}
\end{figure*}

This expression is similar to \eqref{eq:dfRlin}, except the effective ``effective mass" term, $\mu_{\mathrm{eff}}^2$, is now a function of $\tilde{r}$ rather than a constant.
Since $\mu_{\mathrm{eff}}^2(\tilde{r})$ is a smooth function, the solution to this equation can be accurately approximated by slightly modifying the previous Green's functions to include $\mu_{\mathrm{eff}}^2(\tilde{r})$ given above, as 

\bea
\varphi_{G}(\tilde{r},\tilde{R})&=&\frac{8 \pi a G \bar{\rho}}{3 \mu_{\mathrm{eff}}(\tilde{r})} \frac{\tilde{R}}{\tilde{r}} R_{\mathrm{eff}}\times\nonumber
\\
&&\left\{
\begin{array}{lll} 
  e^{-a \tilde{R}R_{\mathrm{eff}} \mu_{\mathrm{eff}}(\tilde{r})} \mathrm{sinh}(a \tilde{r}R_{\mathrm{eff}}\mu_{\mathrm{eff}}(\tilde{r}))
 & \text{, $\tilde{r} \leq \tilde{R}$ } 
 \\   
    {e^{-a \tilde{r}R_{\mathrm{eff}} \mu_{\mathrm{eff}}(\tilde{r})} \mathrm{sinh}(a \tilde{R}R_{\mathrm{eff}}\mu_{\mathrm{eff}}(\tilde{r}))} & \text{, $\tilde{r}>\tilde{R}$} \\
  \end{array}\right.
\label{eq:phiGdefNL}
\eea
with 
\begin{equation}
\varphi_{lin,(i)} (\tilde{r}) \approx \int \mathrm{d}\tilde{R} \left(\epsilon_{(i)}(\tilde{R}) \varphi_{G}(\tilde{r},\tilde{R}) \right).
\end{equation}

After each iterative step, we check to see if $\epsilon$ has converged and, if not, the next iterative step is taken with new trial solution, $ f_{R,(i+1)} = f_{R,(i)}+ w \varphi_{lin,(i)}$ where $w$ is a numerical weight. The default value of $w=1$ is used, except in rare cases in which the initial linear solution  strongly deviates from the nonlinear solution, parametrized by $\vert \epsilon_{(i)} \vert >0.3$, in which we use $w=0.75$ to allow the solution to evolve more conservatively and avoid interative trials ``overshooting" and taking unphysical values.  The iterative procedure is repeated  until $\vert \epsilon_{(i)}(\tilde{r})\vert <0.0075$ over the range $0.2<\tilde{r}<4.8$ (the lower limit avoids numerical ambiguities with the $\nabla^2$ term at $\tilde{r}=0$) after which $f_{R,(i)}$ is considered to have sufficiently converged to the solution the full nonlinear field equation. 

The algorithm is run for each void individually and then the results are averaged. For an individual void, the force profile is calculated using the density contrast from the particles out to $\tilde{r}=5$, at which point the field is taken to be effectively at its background value. 
 
\begin{figure*}[!t]
\includegraphics[width=1\linewidth]{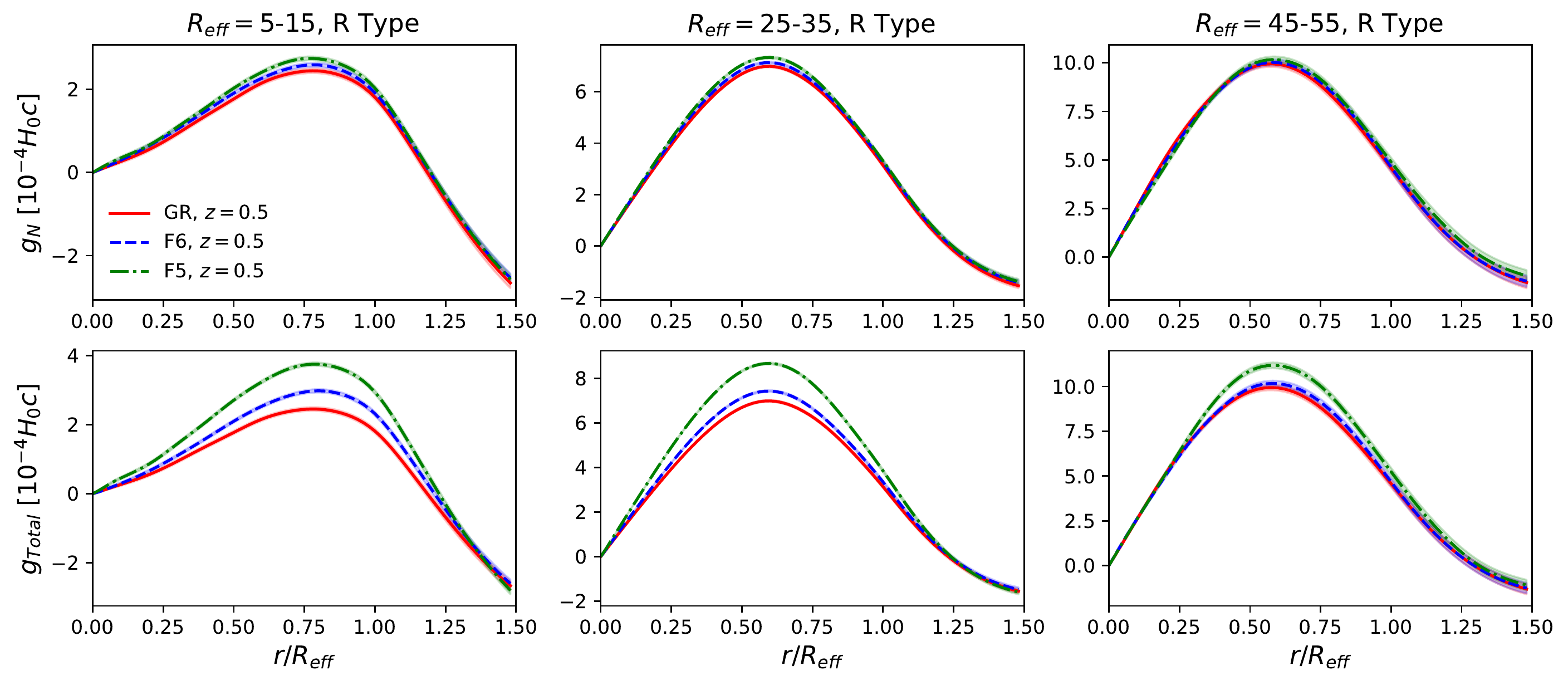}

\caption{The mean Newtonian, $g_N$, [upper panels] and total forces $g_{Total}$ (fifth and Newtonian combined) in GR [full, red line], F6 [blue dashed line] and F5 [green dot-dashed line]  for \Rtype voids for sizes $R_{\mathrm{eff}}=$5-15$Mpc/h$ [left panels], 25-35$Mpc/h$ [center], 45-55$Mpc/h$ [right]. Standard errors on the mean are shown as shaded regions around the mean.}
\label{fig:totalforces}
\end{figure*}

 GR [full, red line], F6 [blue dashed line] and F5 [green dot-dashed line]

Figure~\ref{fig:fifthForces} presents the fifth force per unit mass calculated with the iterative approach for \Rtype voids in F5 and F6 at $z=0$ and $z=0.5$. 
As an indicator of the convergence process, for $F6$ at $z=0.5$,  93\% of \Rtype voids meet the convergence criteria after a single iterative step beyond the linear solution, and $99\%$ of \Rtype voids after three iterations. Only $0.05\%$ of \Rtype voids fail to converge below $\vert \epsilon_{(i)}(\tilde{r}) \vert <0.05$ after ten iterations and were excluded from the force analysis. Although not shown, we found good agreement between the fifth force calculated from the average density profile using the average $R_{\mathrm{eff}}$ within each size bin, and the average of the fifth forces calculated for each density profile individually. 

In $F6$, the magnitude of the peak of the fifth force is a strictly decreasing function of $R_{\mathrm{eff}}$, and acts at smaller $\tilde{r}$ as one goes to larger voids. This is consistent with the characteristics of the velocity profiles in Figs.~\ref{fig:partRadVelz0p0} and \ref{fig:haloRadVelBoth}, and the $F6$ window function peaking at smaller $R_{\mathrm{eff}}$, as shown in Fig.~\ref{fig:WindowFunction}. 
In $F5$, there is far less dependence on void size, consistent with the broad maximum in the $F5$ window function that spans the intermediate size voids in Fig.~\ref{fig:WindowFunction}. 
For completeness, the rightmost panel of Fig.~\ref{fig:fifthForces} shows the Newtonian force in GR, which is a strictly increasing function of void size.

\begin{table}[!b]
\begin{tabular}{|c|c|c|c|c|c|}
\hline
\multirow{2}{*}{Model  }    & \multirow{2}{*}{$z$} & Void Size & \multirow{2}{*}{$g_{5}/g_{N,GR}$ }& \multirow{2}{*}{$g_{N,f(R)}/g_{N,GR}$} & \multirow{2}{*}{$g_{Total, f(R)}/g_{N,GR}$ }
\\ 
 & & ($Mpc/h$) &&& \\
\hline \hline
\multirow{6}{*}{$F6$}	& 	\multirow{3}{*}{$0$}  	  		
		& $5-15$   	& 	$0.15\pm0.004$	& 	$1.04 \pm 0.03$	& $1.17 \pm 0.03$            
\\ \cline{3-6} & 
         & $25-35$  	& 	$0.03 \pm0.0004$	& 	$1.03 \pm 0.01$	& $1.06 \pm 0.01$          
\\ \cline{3-6} & 
         & $45-55$   	& 	$0.01 \pm0.0004$ 	& 	$1.02 \pm 0.02$	& $1.03 \pm 0.02$                    
\\ \cline{2-6} & 	{\multirow{3}{*}{$0.5$}}  
		& $5-15$  		& $0.17 \pm 0.004$        & $1.06 \pm 0.03$        & $1.22 \pm 0.04$           
\\ \cline{3-6} &              
		& $25-35$   			& $0.04 \pm 0.0004$       & $1.02 \pm 0.01$        & $1.06 \pm 0.01$            
\\ \cline{3-6} &
		& $45-55$   			& $0.02 \pm 0.0005$       & $1.01  \pm 0.02$       & $1.02 \pm 0.02$           
\\ \hline \hline \multirow{6}{*}{$F5$}	& 	\multirow{3}{*}{$0$}
		& $5-15$   		& $0.39 \pm 0.009$        & $1.09 \pm 0.03$         & $1.47 \pm 0.04$            
\\ \cline{3-6} &
		& $25-35$   			& $0.17 \pm 0.001$        & $1.05 \pm 0.01$        & $1.22 \pm 0.01$           
\\ \cline{3-6} & 
		& $45-55$   			& $0.08 \pm 0.002$        & $1.04 \pm 0.02$       & $1.12  \pm 0.02$        
\\ \cline{2-6}& \multirow{3}{*}{$0.5$}
		& $5-15$   		& $0.41 \pm 0.01$        & $1.12 \pm 0.04 $       & $1.53  \pm 0.04$          
\\ \cline{3-6} & 
		& $25-35$  			& $0.19 \pm 0.002$       & $1.05  \pm 0.01$       & $1.24 \pm 0.01$           
 \\ \cline{3-6} & 
		& $45-55$  			& $0.10  \pm 0.003$       & $1.02 \pm 0.03$      & $1.12 \pm 0.02$           
\\ \hline
\end{tabular}
\caption{A summary of the mean ratios of the fifth forces, $g_5$ [left columns], Newtonian $g_{N,f(R)}$ [central columns] and total forces, $g_{Total, f(R)}$ [right columns] experienced in \Rtype voids in $f(R)$ relative to the Newtonian forces experienced in GR, $g_{N,GR}$. Results are shown for voids of various sizes, for both F5 and F6 scenarios at $z=0$ and $z=0.5$. Errors are the 1$\sigma$ errors on the mean values.}
\label{tab:forceratios}
\end{table}

Figure~\ref{fig:totalforces} and Table~\ref{tab:forceratios} show the relative importance of the Newtonian force and the total force, including the fifth force for $z=0.5$ voids. The relative strength of the fifth force is largest in the smallest voids both for $F6$ and $F5$. For $F6$ at $z=0.5$, the fifth force is $17\% \pm 0.4\%$ of the Newtonian force in $GR$ at its peak in the smallest voids, while it contributes a significantly smaller fraction, $2\% \pm .05\%$, at the peak in the largest voids. 

The ratio of the Newtonian forces, $g_{N,f(R)}/g_{N,GR}$ differs slightly from unity when comparing $F5$ and $F6$ to $GR$, resulting from small differences in the particle density profiles shown in Fig.~\ref{fig:densitiesAcrossScale} in early $r/R_{\mathrm{eff}}$ bins. In $F5$, the magnitude of the fifth force is much larger than this difference in Newtonian forces. In $F6$, the fifth force is much larger in small voids, while it is more comparable to  the difference in Newtonian forces in the $25-35 Mpc/h$ and $45-55 Mpc/h$ voids. Interestingly, the sets of voids with the largest fractional difference in Newtonian forces are also those with the largest fractional fifth forces, indicating that the fifth force is playing an active role in shaping these environments.

It is frequently stated in the literature that $f(R)$ gravity can provide a maximum enhancement of a factor of $4/3$ over Newtonian gravity in $GR$. This is derived from the ratios of the coupling constants in  (\ref{eq:dfR}) and (\ref{eq:Newton}). In the context of specific matter distributions, however, an  enhancement  greater than $4/3$ can be obtained under the assumption of spherical symmetry.  As an example, if a thin spherical shell of radius $R$ were considered, the ratio $g_{5}/g_{N}$ would be infinite at the points $0<r<R$ interior to the shell without contradicting the theoretical mode, since the Newtonian force within the shell would be zero, while the fifth force will be nonzero, as described in Sec.~\ref{sec:linfifthGreens}. It is in this context that  the values in Table \ref{tab:forceratios} for $g_{Total, f(R)}/g_{N,GR}$ should be understood. The values $> 4/3$ are a result of the assumption of spherical symmetry, combined with small differences in the underlying density profiles used to calculate $g_{N}$ in $GR$ and $g_{Total}$ in $F5$ and $F6$. Assumptions of spherical symmetry have shown to be reasonable and to give results which match those directly from simulations e.g. for nDGP models of gravity \cite{Falck:2017rvl}. Collectively, these results underline why velocity profiles within small \Rtype voids present a robust method to isolate distinctive signatures of modified gravity resulting from the direct action of the fifth force. 

\subsection{The Chameleon Mechanism and Screening Factor, $\alpha$}
\label{sec:screening}

\begin{figure*}[!t]
\includegraphics[width=1\linewidth]{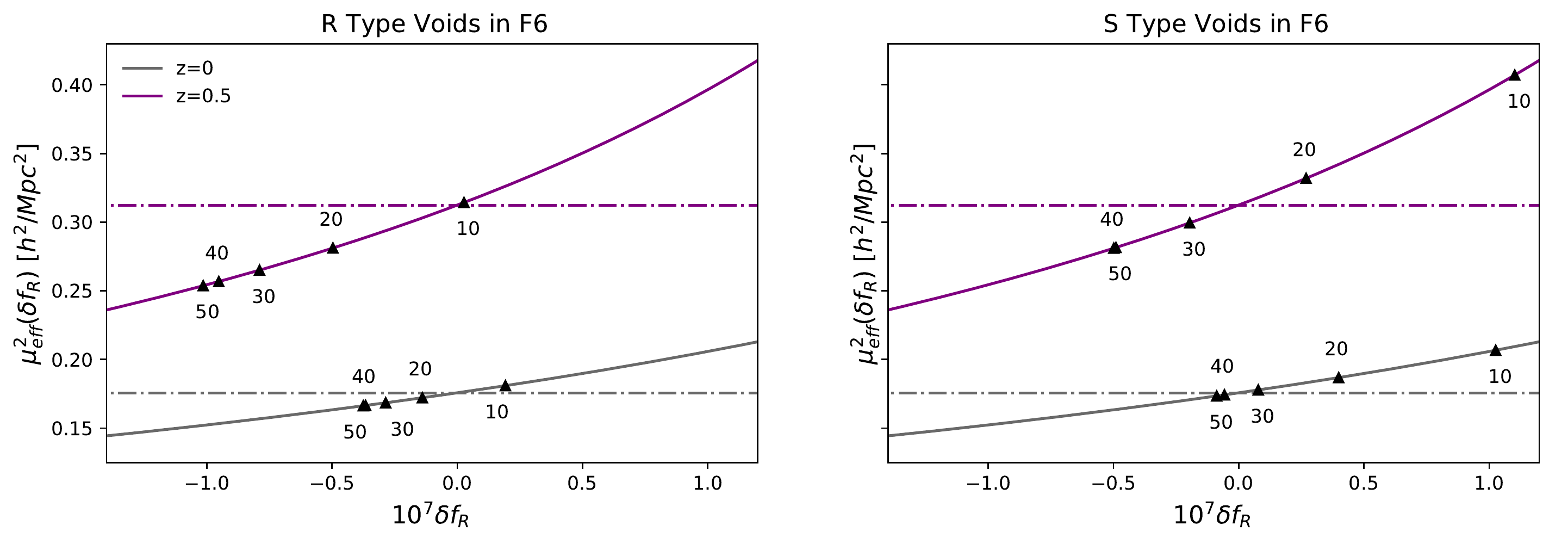}
\caption{The effective mass calculated at the peak of the fifth force, $\mu_{\mathrm{eff}}^2$, in F6 for \Rtype [left panel] and \Stype [right panel] voids. The effective mass estimates obtained analytically using \eqref{eq:muEffExplicit} are shown for $z=0$ [grey full lines] and $z=0.5$ [purple full lines] along with the homogeneous background value $\mu^2(z)$ [dot-dashed lines]. The results from simulated void data, labeled by the $R_{\mathrm{eff}} (Mpc/h)$ of the respective size bins are also shown [triangle markers].}
\label{fig:muEffF6}
\end{figure*}

\begin{figure*}[!t]
\includegraphics[width=1\linewidth]{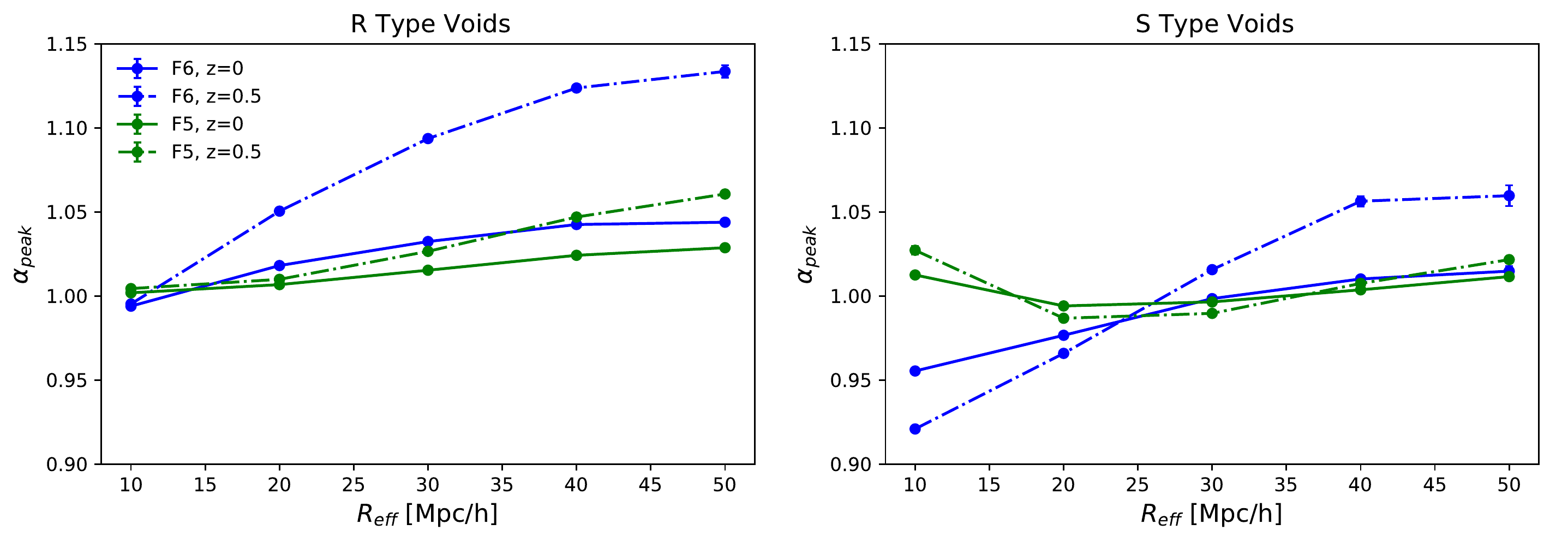}
\caption{The value of the screening factor, $\alpha$, calculated at the peak value of the fifth force. F6 [blue] and F5 [green] models as a function of void size are shown for the \Rtype [left] and \Stype voids [right] at $z=0$ [full lines] and $z=0.5$ [dot dashed lines].}
\label{fig:alpha}
\end{figure*}

The quality of the match between the solutions of the linearized and the full nonlinear field equation is encapsulated by the screening factor $\alpha$ in \eqref{eq:alpha}. Before we present our results for the screening factors, it is instructive to first consider the effective mass previously defined in  (\ref{eq:muEff}).  The effective mass, $\mu_{\mathrm{eff}}^2$, is a useful measure of the total amount of screening at play, whereas the screening factors will only inform us to additional screening or enhancement on top of the linear fifth force solutions. Here, the effective mass can be written in a more explicit way, separating the field $f_{R}$ into $f_{R}=\bar{f}_R+ \delta f_{R}$:
\begin{equation}
\mu_{\mathrm{eff}}^2(\tilde{r})= - \frac{\bar{R}_{0}}{3(n+1) \left(\bar{f}_{R} + \delta f_{R}(\tilde{r}) \right)}\left(\frac{\bar{f}_{R,0}}{\bar{f}_{R} + \delta f_{R}(\tilde{r})} \right)^{\frac{1}{n+1}}
\label{eq:muEffExplicit}
\end{equation}
where $\bar{f}_R$ is explicitly given in (\ref{eq:barPlusDelta}). 

In Figure~\ref{fig:muEffF6}, the average value of $\mu^{2}_{\mathrm{eff}}$ is shown, evaluated at the location of the peak fifth force within each void in the $F6$ simulations, and averaged over all voids within the same classification and size bin. $\mu_{\mathrm{eff}}^2$ calculated in this way is found to be a decreasing function of $R_{\mathrm{eff}}$, and decreases more sharply at a redshift of $z=0.5$ than at $z=0$. It is consistently smaller in \Rtype voids compared to the similarly sized \Stype voids. 

Examining \eqref{eq:dfR}, due to the sign of the matter coupling, regions of $\delta>0$ act to drive $\delta f_{R}$ positive, closer to its fully screened value of $8 \pi G  \bar{\rho} \delta$ and likewise regions of $\delta<0$ act to drive $\delta f_{R}$ negative. As per the chameleon mechanism, looking at \eqref{eq:muEffExplicit}, overdense regions drive $\delta f_{R}$ positive, causing the entire field $\bar{f}_{R} +\delta f_{R}$ to grow smaller in magnitude, and thereby increasing the effective mass over from its background value.  \Rtype voids feature a smaller overdense shell than the \Stype voids, as shown in Figure~\ref{fig:densitiesAcrossScale}, and thus a smaller accompanying value of $\mu_{\mathrm{eff}}$.

The greater variation in $\delta f_R$ at $z=0.5$ relative to $z=0$ can be understood by noting that $a^2 \mu(a)^2$, the combination of which acts as the linear mass term in (\ref{eq:dfRlin}), is smaller at $z=0.5$ than $z=0$ due to the explicit inclusion of the scale factor. Thus, the $\delta f_{R}$ field typically acquires larger values at $z=0.5$ compared to $z=0$, and thus more extreme values of $\mu_{\mathrm{eff}}^{2}$. This trend will not continue to earlier redshifts as $a^2 \mu(a)^2$ is minimized for $z\simeq 0.4$ and is a strictly increasing function with increasing redshift beyond this point. 

Figure~\ref{fig:alpha} shows how $\alpha$, when evaluated at the location of the peak outwards fifth force, varies as a function of void size within \Rtype and \Stype voids for both $F5$ and $F6$. The figure shows that in voids, the linear solutions provide a reasonable but not exact solution to the nonlinear equations (when they agree perfectly, $\alpha=1$).

As one moves to lower redshifts, the linearized field in \eqref{eq:dfRlin} provides an increasingly accurate approximation to the full field equation within void environments of all classifications and sizes. This can be explained using  \eqref{eq:barPlusDelta} and \eqref{eq:muEffExplicit}. Equation (\ref{eq:muEffExplicit}) shows how the nonlinear effects, which act to greatly increase the field's mass kick in when $\delta f_{R} \sim \vert \bar{f}_{R} \vert$, cause $f_{R}$ to approach $0$ from below. In (\ref{eq:barPlusDelta}), one can see that as $a \rightarrow 0$, $\vert \bar{f}_R \vert $ grows larger in magnitude, and $\delta f_{R}$ is allowed to operate over a larger range of values before nonlinear effects become significant. Related is the fact that as $\bar{f}_R$ grows larger in magnitude, the $\mu_{\mathrm{eff}}$ curves in Figure \ref{fig:muEffF6} flatten out; for larger values of $\vert \bar{f}_R \vert$, changes in $\delta f_{R}$ cause smaller changes to $\mu_{\mathrm{eff}}$. Thus, as $z \rightarrow 0$, it becomes harder to trigger additional screening or enhancement, meaning the linear equation, which contains neither of these effects, becomes a better approximation.

The average value of $\alpha$ calculated at the peak of the fifth force is a monotonically increasing function of $R_{\mathrm{eff}}$ with the exception of small \Stype voids in $F5$. Ignoring this exception for the time being, the monotonic trend is a reexpression of that first observed in Fig.~\ref{fig:muEffF6}. Larger voids have smaller overdense shells, and thus the underdense centers can provide more nonlinear enhancement through the chameleon mechanism compared to their small counterparts. When viewing these figures, one must keep in mind that the value of $\alpha$ for a given void class, radius, and redshift conveys the fractional change from the screening that is already accounted for in the linearized fifth force equation, rather than the \textit{total} amount of screening at play, indicated by $\mu_{\mathrm{eff}}^2$.

The exception to the trends in $\alpha$ for small (5-15$Mpc/h$) \Stype voids in $F5$ can be traced back to the explicit violation of the shell theorem by the fifth force in the $\tilde{r}< \tilde{R}$ branch of (\ref{eq:fifthFG}), which is where the dominant fraction of the outward fifth force in these voids originates. In the limit that $\mu \rightarrow 0$, the shell theorem is restored, whereas in the opposite limit of $\mu \rightarrow \infty$, the field is infinitely massive and cannot propagate. Both limits lead to the same result in (\ref{eq:fifthFG}) of $\mathcal{F}_{5}(\tilde{r},\tilde{R}) \rightarrow 0$ for $\tilde{r}< \tilde{R}$, with an intermediate value of $\mu$ maximizing the average fifth force interior to any mass shell. Analogous to the discussion in Sec.~\ref{sec:linfifthGreens}, if one integrates over $\tilde{r}< \tilde{R}$, we find the average fifth force inside a matter shell located at $\tilde{R}$ to be maximized for $\mu_{max} \simeq {2.73}/{a \tilde{R} R_{\mathrm{eff}}}$. If we consider values specific to the 5-15$Mpc/h$ \Stype voids in $F5$ at $z=0.5$, with $\tilde{R}\sim 1.0$ (the location of the peak overdensity in small \Stype voids) and $R_{\mathrm{eff}} \sim12.1 Mpc/h$ (the mean \Stype void size in the smallest bin), we find $\mu_{max} \sim 0.34 h/Mpc$. This is larger than the $z=0.5$ background values of $\mu=0.177 Mpc/h$ in $F5$. Thus, increasing $\mu_{\mathrm{eff}}$ from its background value will have the effect of increasing the outward fifth force in the smallest \Stype voids in $F5$. This is exactly what the chameleon mechanism does, increasing the mass of the field at the peak of the fifth force within these voids on average by $13\%$ up to $\mu=0.20 h/Mpc$. The story breaks down and reverts to the more intuitive case for \Rtype voids and the larger \Stype voids in F5, with most of their fifth force coming instead from the $\tilde{r}>{R}$ branch of (\ref{eq:fifthFG}). For both types of voids in F6, the linear mass term at $z=0.5$ of $\mu=0.56 h/Mpc$ at $z=0.5$, is much greater than the corresponding $\mu_{max}$ in all sizes of voids; further increases to $\mu_{\mathrm{eff}}$ will only dampen the fifth force in $F6$ voids. 

It is instructive to compare the screening factor we have obtained in the voids to the approximate  screening factor screening  for spherically compact objects proposed by Khoury and Weltman \cite{Khoury:2003aq,Khoury:2003rn}  for large, spherically overdense objects of radius $R_{obj}$ in $f(R)$ gravity, as  
\bea
 \left(\frac{\Delta R_{obj}}{R_{obj}}\right)=  \begin{cases} 
    \mathrm{min}\left(\frac{3}{2}  \lvert\frac{\bar{f}_{R_0}}{\Phi_N}\rvert \left( \frac{\Omega_{m0}+4\Omega_{\Lambda 0}}{\Omega_{m0}a^3+4\Omega_{\Lambda 0}}\right)^{n+1},1\right)
 & \text{if $\Phi_N < 0$ } \\
      
       1 & \text{if $\Phi_N \geq 0$ }. \\
   \end{cases}
   \label{eq:denseScreen}
\eea

Comparing our void screening factor $\alpha$ against this, we find substantial differences. Most notably, whereas the void screening factor $\alpha$ allows for enhancement of the fifth force beyond the linearized value, no such enhancement is allowed when using \eqref{eq:denseScreen} in the case of dense objects. There is also a difference in how the two screening factors treat $F5$ versus $F6$. With the void screening factor $\alpha$, \Rtype voids in $F6$ consistently receive a larger fractional enhancement over their linearized fifth force values than their $F5$ counterparts, whereas in \eqref{eq:denseScreen}, due to the explicit inclusion of $\vert \bar{f}_{R_0} \vert$,  dense objects are screened a factor of 10 more heavily in $F6$ than in $F5$. 

Using the average density profile in each of our $R_{\mathrm{eff}}$ size bins as explicit examples, we can calculate both $\alpha$ and  $\left({\Delta R_{obj}}/{R_{obj}}\right)$ at the location of each profile's peak fifth force and compare. 

Considering \Rtype voids at $z=0.5$, we find the average profile of the 5-15 $Mpc/h$ size bin to have $\alpha=0.98$ whereas the dense object screening factor gives a markedly different answer of $\left(\frac{\Delta R_{obj}}{R_{obj}}\right)=0.21$. For 15-25$Mpc/h$ and 25-35$Mpc/h$, we find $\alpha=1.07$ and $\alpha=1.14$ whereas $\left(\frac{\Delta R_{obj}}{R_{obj}}\right)=0.83$ and $\left(\frac{\Delta R_{obj}}{R_{obj}}\right)=0.80$ for each size range respectively, indicating that while the forces are actually enhanced over their linearized values, the dense object screening factor would add additional nonlinear screening. In the larger size bins, we have $\left(\frac{\Delta R_{obj}}{R_{obj}}\right)=1$, while in each case $\alpha$ takes a value greater than one, $\alpha=1.16$, and $\alpha=1.18$ in the 35-45$Mpc/h$ and 45-55 $Mpc/h$ bins respectively.

These results have implications for studying voids in $f(R)$ gravity using hybrid simulation techniques, that combine N-body and Lagrangian perturbation theory approaches \citep{Winther_2017, Valogiannis_2017}. These implement the Chameleon mechanism through the use of the compact object screening factor $\left({\Delta R_{obj}}/{R_{obj}}\right)$ and have been shown to create clustering statistics that agree well with results with full N-body simulations which solve the nonlinear field equations. These statistics, however,  principally focus on  regions of high density. Our work provides an approach to be able to extend these hybrid approaches to the study of voids, by analytically calculating $\alpha$ using iterative method developed here  to solve the full nonlinear field equation for the fifth force, in regions where the compact object form does not apply.

\section{Conclusions}
\label{sec:conc}

In this paper, we determine how halo velocities within voids can be used to discriminate between GR and $f(R)$ gravity by contrasting void velocity profiles across classifications and a range of void sizes. 

Voids are identified in snapshots from  N-body simulations at $z=0$ and $z=0.5$ using the void finder VIDE and are classified based on their halo density profiles as either \Rtype (rising) or \Stype (shell), and analyzed in groups based on their effective radius, $R_{\mathrm{eff}}$. We find few observable differences in the halo-derived density profiles in voids of either classification or size, although when dark matter particles are  used as tracers, we find slightly emptier voids at small $r/R_{\mathrm{eff}}$ in modified gravity scenarios consistent with previous work \cite{Zivick:2014uva}. 

We find that the velocity profiles of \Rtype voids in modified gravity scenarios are much more distinguishable from their $GR$ counterparts than for \Stype voids. This effect is most pronounced in the smallest with voids $5(Mpc/h)<R_{\mathrm{eff}}<15(Mpc/h)$, which provide the best dynamical opportunity to distinguish between $F6$, the most weakly modified gravity  scenario considered, and $GR$.  The difference in velocity profiles is observed in both the halo and particle velocity profiles, and at $z=0.5$ and $z=0$. The peak velocities in these voids, using the halo data, is found to be $13\%\pm8\%$ larger at $z=0$ and $8\%\pm6\%$ larger at $z=0.5$ in $F6$, and $28\%\pm8\%$ at $z=0$ and $22\%\pm7\%$ at $z=0.5$ in $F5$ when compared to $GR$.

We undertake a detailed analysis of the fifth and Newtonian forces and are able to attribute the signal in the small voids to the action of the fifth force as opposed to underlying differences in void populations or density profiles across the simulations. The analysis of the linearized field equation through the use of the window functions shows that the linearized fifth force in $F6$ will be a decreasing function of void size, whereas in $F5$ there will be much less size dependence on the magnitude of the linearized fifth force. The ratio of the linearized fifth force in either modified gravity scenario to that of the Newtonian force is shown to be maximized in small voids. 

We develop an iterative procedure, using Green's functions, to solve the nonlinear field equation in voids under the assumption of spherical symmetry. The method efficiently enables the fifth force to be calculated in each void individually, rather than just for the mean density profile. 

Comparing the linear and full solution to the field equation, we compute the screening factor $\alpha$. We find that in all voids, $\alpha$ is of order unity, but differs from unity depending on the size and void classification in both $F5$ and $F6$. The screening factor $\alpha$ is found to be consistently larger (meaning less screening is occurring) in \Rtype voids compared to \Stype voids, and large voids compared to small voids. The value of $\alpha$ is more easily displaced from unity in either direction at $z=0.5$ compared to $z=0$, indicating that nonlinear effects are more important at earlier redshifts. 

Focusing on $F6$, we can see there is competition between the screening or enhancement to the fifth force given by $\alpha$, which is found to increase with $R_{\mathrm{eff}}$, and the linearized fifth force analysis, which states that the magnitude of the fifth force should decrease with increasing $R_{\mathrm{eff}}$. Considering \Rtype voids in $F6$ at $z=0.5$, we find that despite a larger average screening factor of $\alpha=1.13$ in voids with $R_{\mathrm{eff}}=45-55 Mpc/h$, the largest fifth force is found to be in the smallest voids with $R_{\mathrm{eff}}=5-15 Mpc/h$, which an average screening factor value of $\alpha=1.00$. This result shows that ultimately, the linear force analysis dictates the trends which occur in the full nonlinear fifth force with changing $R_{\mathrm{eff}}$.

We also find screening in voids cannot be effectively captured using the widely-used screening factor approximation developed for spherically overdense bodies in \cite{Khoury:2003rn}, and given explicitly in \eqref{eq:denseScreen}. Considering $F6$ again at $z=0.5$, we find severe mismatch between the values of $\alpha$ and $\left({\Delta R_{obj}}/{R_{obj}}\right)$ both calculated for the same density profile at the location of the peak fifth force -- with the worst discrepancy coming in the voids with the largest distinguishing velocity signal. Given these actual discrepancies, as well theoretical concerns around the lack of potential enhancement to gravity when using the $\left({\Delta R_{obj}}/{R_{obj}}\right)$ screening factor, we discourage the use of hybrid codes which implement this screening factor when studying cosmic voids, and instead encourage the use of alternative methods.

Our results present tantalizing prospects for constraining the properties of gravity through looking at void statistics with redshift space distortion measurements from DESI, Euclid and the Roman Telescope. Photometric surveys, such as the Rubin Observatory LSST survey, will also provide additional valuable information to accurately determine the density profiles that aid the characterization of void sizes and classifications. Determining the full observational implications for upcoming large scale structure surveys will be the focus of future work. 

\section*{Acknowledgements}
We wish to thank Baojiu Li for kindly providing the ELEPHANT simulations, on behalf of \citep{Cautun:2017tkc} and Georgios Valogiannis for assistance in their use. The work of Christopher Wilson and Rachel Bean is supported by DoE grant DE-SC0011838, NASA ATP grant 80NSSC18K0695, NASA ROSES grant 12-EUCLID12-0004 and funding related to the Roman High Latitude Survey Science Investigation Team. 

\newpage
\bibliographystyle{apsrev}
\bibliography{references}

\end{document}